\documentclass[english,11pt,a4paper,DIV=16,numbers=noenddot]{scrartcl}
\pdfoutput=1
\usepackage{axodraw}
\usepackage[english]{babel}
\usepackage{cite}
\usepackage{a4wide}
\usepackage{amsmath}
\usepackage[charter,cal=cmcal,greekuppercase=italicized]{mathdesign}
\usepackage{dsfont}
\usepackage{physics}
\usepackage[T1]{fontenc}
\usepackage{slashed}
\usepackage{setspace}
\usepackage{graphicx}
\usepackage{xcolor}
\usepackage{booktabs}
\usepackage{soul}
\usepackage{ulem}
\usepackage{bm}
\usepackage{siunitx}
\usepackage[final,          
            colorlinks,     
            linkcolor=purple, 
            citecolor=teal, 
            urlcolor=blue  
            ]{hyperref}

\usepackage[small,nooneline]{caption}
\usepackage{subcaption}
\usepackage{afterpage}
\makeatletter
\DeclareRobustCommand*{\bfseries}{%
  \not@math@alphabet\bfseries\mathbf
  \fontseries\bfdefault\selectfont
  \boldmath
}
\makeatother

\newcommand{\el}{\ensuremath{\mathrm{e}}}

\newcommand{\im}{\ensuremath{\mathrm{i}\,}}

\newcommand{\Lag}{\ensuremath{\mathcal{L}}}
\newcommand{\ie}{i.\,e.~}

\newcommand{\Braket}[2]{\ensuremath{\left\langle#1\right|\left.\!#2\right\rangle}}

\hypersetup{
  pdfauthor={Wolfgang G. Hollik, Matthias Linster, Mustafa Tabet}
  ,pdftitle={A Study of New Physics Searches with Tritium and Similar Molecules}
  ,pdfsubject={}
  ,pdfkeywords={}
}

\title{\vspace*{-3cm}
  \begin{flushright}
    {\textsf{\small
     TTP-2020-016, P3H-20-014}\\
    }
  \end{flushright}
  \vfill
  A Study of New Physics Searches with Tritium and Similar Molecules}

\author{Wolfgang Gregor Hollik${}^{a,b}$,
  Matthias Linster${}^{b}$,
  Mustafa Tabet${}^{b}$
  \\[1em]
\textit{${}^a$Institute for Nuclear Physics, Karlsruhe Institute of Technology,}\\
\textit{D-76021 Karlsruhe, Germany}\\[.7em]
\textit{${}^b$Institute for Theoretical Particle Physics, Karlsruhe Institute of Technology,}\\
\textit{D-76128 Karlsruhe, Germany}\\[.7em]
  \\[1.5em]
{\small\texttt{wolfgang.hollik@kit.edu}}\\[-0.3em]
{\small\texttt{matthias.linster@kit.edu}}\\[-0.3em]
{\small\texttt{mustafa.tabet@kit.edu}}
}
\date{\today}

\addtokomafont{title}{\rmfamily}
\addtokomafont{section}{\rmfamily}
\addtokomafont{subsection}{\rmfamily}

\begin{document}
\maketitle
\thispagestyle{empty}

\begin{abstract}
\noindent
Searches for New Physics focus either on the direct production of new
particles at colliders or at deviations from known observables at low
energies.  In order to discover New Physics in precision measurements,
both experimental and theoretical uncertainties must be under full
control.  Laser spectroscopy nowadays offers a tool to measure
transition frequencies very precisely. For certain molecular and
atomic transitions the experimental technique permits a clean study of
possible deviations. Theoretical progress in recent years allows us to
compare \emph{ab initio} calculations with experimental data. We study
the impact of a variety of New Physics scenarios on these observables
and derive novel constraints on many popular generic Standard Model
extensions. As a result, we find that molecular spectroscopy is not
competitive with atomic spectroscopy and neutron scattering to probe
new electron-nucleus and nucleus-nucleus interactions,
respectively. Molecular and atomic spectroscopy give similar bounds on
new electron-electron couplings, for which, however, stronger bounds
can be derived from the magnetic moment of the electron. In most of
the parameter space H\(_2\) molecules give stronger constraints than
T\(_2\) or other isotopologues.

\end{abstract}

\newpage

\tableofcontents\label{TOC}
\begin{center}
\hrule
\end{center}

\onehalfspacing
\section{Introduction}\label{sec:intro}
So far, no heavy new particles beyond those of the Standard Model of
elementary particle physics have been found. There is, however,
observational evidence of physics not covered in the Standard Model. In
the context of Dark Matter, for example, the interest to search for new
sub-\si{\giga\electronvolt} particles has recently gained
impetus~\cite{Knapen:2017xzo, Dror:2019onn, Dror:2019dib}, where
molecules have been identified as good study
objects~\cite{Arvanitaki:2017nhi}.  In particular, molecular
spectroscopy is one possibility to look for new dark
forces~\cite{Fichet:2017bng, Brax:2017xho}.  Although the laws of
Quantum Mechanics as the physical framework at molecular scales are well
established, it is intrinsically difficult to provide reliable precise
predictions.  Theoretical calculations are challenging, CPU-intensive,
and potentially lacking important higher-order contributions that might
have been neglected. Nevertheless, there are precise state-of-the art
predictions for hydrogen-like molecules that can be exploited for a
dedicated analysis of New Physics effects.

Following the early groundbreaking works of Ko\l{}os and Wolniewicz in
the 1960s~\cite{Kolos:1962, Kolos:1963, Kolos:1964adiab,
  Kolos:1964vibro, Kolos:1968impr}, a vast progress in the theoretical
determination of energy levels of hydrogen-like molecules has been
made during the last decade.  The crucial improvement was a clear
conceptional separation of electronic and nuclear motion developed in
form of nonadiabatic perturbation theory by Pachucki and
Komasa~\cite{Pachucki2008, Pachucki:2009xy, Pachucki:2010pccp,
  Pachucki:2015xy}. With this method, a full nonadiabatic treatment of
the system is performed. Furthermore,
leading~\cite{Puchalski:2018pdf,Puchalski:2019prl, Puchalski:2019pra}
and higher-order~\cite{Puchalski:2016xy} QED corrections have been
implemented, as well as relativistic
corrections~\cite{Puchalski:2017pra, Puchalski:2018xy}.  Precise
theoretical predictions of rovibrational lines for hydrogen
isotopologues have been made publicly available in the computer code
\texttt{H2Spectre}~\cite{H2spectre, czachothes} and are nicely
reviewed in Reference~\cite{Komasa:2019}.

On the experimental side, improved techniques allow for more precise
measurements of these spectra, testing the theoretical prediction to
unprecedented accuracy.  As a consequence of the agreement between
theory and experiment, severe constraints can be put on any deviation
of known physics~\cite{Salumbides:2013aga, Salumbides:2013dua,
  Salumbides:2015qwa}.

The simplest modification of the well-known Coulomb potential
$V_\mathrm{C}$ for the interaction between two point charges
\(q_{1,2}\) which accounts for light New Physics is given by the
addition of a Yukawa-type potential $V_\mathrm{Y}$,
\begin{equation}\label{eq:potdev}
  V_\mathrm{int} (r) = V_\mathrm{C} (r) + V_\mathrm{Y} (r) =
  \frac{\alpha_\mathrm{em}}{r} \left( q_1 q_2 +
    \frac{g_\text{NP}}{4\pi\alpha_\mathrm{em}} \exp(-m r)\right)\,,
\end{equation}
where \(\alpha_\mathrm{em}\simeq 1/137\) is the electromagnetic fine
structure constant. In this way, a new light particle coupling to
known physics leads to a potential which is exponentially suppressed
by this particle's mass \( m \) and is proportional to the coupling
strength \(g_\text{NP}\). Note that the coupling strength \(
g_\mathrm{NP} \) might have both signs depending on the interacting
particles.  A brief comparison of units yields a mass of
\(\order{\si{\kilo\electronvolt}}\) for a typical bond length \(\sim
\SI{1}{\angstrom}\) of a light molecule.  For larger masses the
exponential term in \(V_\mathrm{Y}\) drops too fast to have an effect
on molecular distances, while for lower masses the Yukawa term is too
long-ranged to be distinct from a Coulomb potential and, hence,
redefines \(\alpha_\mathrm{em}\) by a constant
shift~\cite{Jaeckel:2010xx}.

In the simplest case, a Yukawa-like potential as shown in
Equation~\eqref{eq:potdev} might originate from a light scalar
exchange between two bound fermions but may also appear as the leading
contribution from spin-dependent potentials~\cite{Fadeev:2018rfl,
  Costantino:2019ixl}.  Many models include such light particles as
carriers for weak long-range forces, for instance an additional light
Higgs Boson as a scalar mediator~\cite{Silveira:1985rk,
  McDonald:1993ex}, axions~\cite{Moody:1984ba} and axion-like
particles (ALPs)~\cite{Jaeckel:2010ni} as examples for a pseudoscalar
exchange, or Dark Photons as a vector particle~\cite{Holdom:1985ag,
  Fayet:1990wx, Jaeckel:2010xx}.

The low-energy regime has already been explored in other experiments,
for a review see Reference~\cite{Essig:2013lka}.  Precision QED tests
can be performed with atomic spectroscopy, for example in highly
excited Rydberg atoms or isotope shifts in singly ionized divalent
elements~\cite{Delaunay:2017dku, Berengut:2017zuo, Jones:2019qny}.
While these measurements give slightly better constraints than
molecular spectroscopy, long-distance inter-nuclear interactions can
only be tested in molecules---although neutron scattering might be
more competitive~\cite{Kamiya:2015eva}.  Additionally, atomic
precision tests for light scalar couplings have been
considered~\cite{Brax:2010gp}, where light ALPs modify the Coulomb
potential by a screening effect and may have impact on the Lamb Shift
in atomic hydrogen~\cite{Villalba-Chavez:2018eql}.  However, there are
competing laboratory techniques with higher sensitivity as pointed out
in Reference~\cite{Villalba-Chavez:2018eql} and by further dedicated
studies on atomic spectroscopy~\cite{Delaunay:2017dku,
  Berengut:2017zuo, Jones:2019qny}. Especially in the mass regime
below several \si{\mega\electronvolt}, there are stringent indirect
constraints from astrophysics~\cite{Raffelt:1990yz, Raffelt:1996wa,
  Raffelt:2006cw} and cosmology~\cite{Depta:2020wmr}. With atomic and
molecular spectroscopy, new forces at the \si{\kilo\electronvolt}
scale can be probed directly by the single-particle interaction in
contrast to multi-particle coherent effects in massive objects.

Alternatively, the New Physics contribution might be interpreted as a
test of gravity and/or deviations from known gravitational
interactions, often called ``\emph{fifth forces}'' in the
literature. Such \emph{fifth forces} may have different origin like
extra dimensions where the mass parameter $m$ corresponds to the size
of the extra dimension.  A tight upper bound on the New Physics
coupling of \(g_\text{NP} \lesssim 10^{-34}\) can be derived from test
of the gravitational inverse-square
law~\cite{Carugno:1996uc,Wagner:2012ui,Klimchitskaya:2015kxa,
  Klimchitskaya:2017cnn}. However, these contraints apply only to
mediator masses up to the \si{\milli\electronvolt} regime. In that
respect, atomic and molecular spectroscopy are highly relevant as
probes of new forces in the \si{\angstrom}-regime corresponding to
masses of \( \order{\si{\kilo\electronvolt}} \). Such constraints on
\emph{fifth forces} from molecular spectroscopy have been derived for
a Yukawa-like interaction between nuclei in
References~\cite{Salumbides:2013aga, Salumbides:2013dua}.  Similarly,
\emph{fifth force} experiments lead to constraints on light scalars
coupled to photons~\cite{Dupays:2006dp}.

In this work, we study the impact of New Physics potentials on
molecular spectroscopy of the hydrogen isotopologues H\(_2\), D\(_2\),
T\(_2\), HD, DT, and HT. First, we briefly review the current status
of molecular spectroscopy from a theoretical and experimental
perspective in Section~\ref{sec:background}.  Next, we show the
constraints resulting from each type of new interaction in
Section~\ref{sec:results}. Finally, we conclude in
Section~\ref{sec:conclusion}.

\section{Spectroscopy of Molecular Hydrogen and its Isotopologues}
\label{sec:background}

Atomic and molecular spectroscopy have become fields of research at
the precision frontier. Unlike atoms, diatomic molecules contain a
second nucleus leading to vibrational and rotational excitations of
the whole molecule. Therefore, the spectral lines present in molecules
are rather bands comprising many single lines characterized by
vibrational and rotational quantum numbers \( v \) and \( J \),
respectively. The distance of the individual lines within one band is
much smaller than spectral lines of electronic transitions.  As a
consequence, molecular spectra have a richer structure and are
sensitive to phenomena at much smaller energies compared to atomic
spectra.
  
In particular, the accurate determination of rovibrational transitions
probes the internuclear potential and new interactions among electrons
and nuclei.  These transitions are classified by the change \(\Deltaup
J\) in the angular momentum quantum number \( J \) for a certain
vibrational transition, where \( Q \), \( R \), and \( S \) branches
refer to \( \Deltaup J = 0, 1, 2 \), respectively. Thus, the
vibrational transition can be determined with increased statistical
significance by probing such a branch for different values of the
angular momentum \( J \).  This is especially advantageous for New
Physics searches where effects are rotationally invariant so that
lines for different angular momentum values comprise the same New
Physics effect.

Molecular hydrogen and its isotopologues have been thoroughly studied
nowadays. For instance, the fundamental vibrational line corresponding
to the \( v = 1 \rightarrow 0 \) transition has been observed in the
\( Q \) branch, \ie{} in \(\Deltaup J = 0\) transitions, with high
accuracy for H\(_2\), HD, and D\(_2\)~\cite{Dickenson:2013, NIU201444,
  PhysRevLett.120.153002}.  Moreover, the world's best spectra of
molecules containing Tritium have been recently obtained using
Coherent Anti-Stokes Raman Scattering Spectroscopy (CARS) for T\(_2\)
and DT~\cite{Schlosser_2017, Trivikram:2018, Lai2019DT}.  A relative
precision of up to \( \order{\num{e-10}} \) has been reached in these
measurements. Remarkably, theory predictions are able to match the
experimental sensitivity although becoming more complex.

In the following, we briefly review the current status of theory
calculations in Section~\ref{subsec:theoryReview} and of experimental
measurements in Section~\ref{subsec:experimentalStatus}.

\subsection{A Brief Review of the Current Theoretical Status} 
\label{subsec:theoryReview}

A full theoretical treatment of the Hydrogen molecule H\(_2\) as a
four-particle system is intrinsically difficult. First approaches date
back to the 1920s and have been developed independently by Heitler and
London~\cite{Heitler:1927}, and Born and Oppenheimer~\cite{Born:1927}.
The key part of the Born--Oppenheimer approximation is that it is a
formal expansion in the small ratio of electron over nucleus mass in
powers of the \emph{4th root} \(\sqrt[4]{m_\mathrm{e}/m_\mathrm{N}}\),
while Heitler and London neglected the motion of the nuclei in the
Hamiltonian. This effect can be included in the adiabatic
approximation using perturbation theory~\cite{Born:1951}. A consequent
\emph{non-adiabatic} treatment takes the movement of the nuclei into
account in order to calculate the energy levels of the whole
system~\cite{Kolos:1963}.

Assuming the nuclei to be at fixed positions $\vb{R}_\mathrm{A}$ and
$\vb{R}_\mathrm{B}$, the Hamiltonian for this system
reads~\cite{Heitler:1927}
\begin{equation} \label{eq:HeitlerLondon}
  H_\mathrm{el} = \frac{\vb{P}_1^2}{2m_\mathrm{e}} + \frac{\vb{P}_2^2}{2m_\mathrm{e}}
   + \alpha_\mathrm{em} \left\{ - \frac{1}{r_{1\mathrm{A}}} - \frac{1}{r_{2\mathrm{B}}} +
   \left( \frac{1}{r_{12}} + \frac{1}{R_\mathrm{AB}} -
   \frac{1}{r_{1\mathrm{B}}} - \frac{1}{r_{2\mathrm{A}}} \right) \right\} \,,
\end{equation}
with the electromagnetic fine structure constant \(\alpha_\mathrm{em}
\simeq 1/137 \) and the distances \(r_{12}\) and \( R_\mathrm{AB} \)
between the two electrons \(1\) and \(2\) and nuclei A and B,
respectively.  Correspondingly, \(r_{iX}\) denotes the separation of
electron \(i\) from nucleus \(X\), with \(i = 1,2\) and \(X =
\mathrm{A},\mathrm{B}\). The Schr\"odinger equation for this
Hamiltonian~\eqref{eq:HeitlerLondon} is usually solved using a
variational method with a trial wave function \(
\psi_\mathrm{el}(\vb{r}_1, \vb{r}_2; \vb{R}_\mathrm{A},
\vb{R}_\mathrm{B}) \) expanded in a suitable basis. In the case of the
hydrogen ground state, precise results can be obtained using the
symmetric James--Coolidge
basis~\cite{doi:10.1063/1.1749252,PhysRevA.82.032509},
\begin{multline}
  \psi_\mathrm{el}(\vb{r}_1, \vb{r}_2; \vb{R}_\mathrm{A}, \vb{R}_\mathrm{B}) = \hat{S} \sum_{n_0, n_1, n_2, n_3, n_4} C_{n_0,\dots,n_4} R_\mathrm{AB}^{-3 - n_0 - n_1 - n_2 - n_3 - n_4} \: \mathrm{e}^{-u(r_{1\mathrm{A}} + r_{1\mathrm{B}} + r_{2\mathrm{A}} + r_{2\mathrm{B}})} \\
 \times  r_{12}^{n_0} (r_{1\mathrm{A}} - r_{1\mathrm{B}})^{n_1} (r_{2\mathrm{A}} - r_{2\mathrm{B}})^{n_2} (r_{1\mathrm{A}} + r_{1\mathrm{B}})^{n_3} (r_{2\mathrm{A}} + r_{2\mathrm{B}})^{n_4}
\label{eq:James-Coolidge basis}
\end{multline}
with variational parameter \( u \), non-negative integers \( n_i\),
\(i = 0, 1, \dots, 4 \), and the symmetrization operator $\hat{S}$ to
satisfy the Pauli principle.

Now, the effects of the nuclear motion and kinetic interaction between
electrons and nuclei are described by the Hamiltonian
\begin{equation}\label{eq:nonadH}
  H_\mathrm{n} = - \frac{1}{2\mu_\mathrm{n}} \left(\laplacian_R +
    \laplacian_\text{el} \right) + \left(\frac{1}{M_\mathrm{A}} -
    \frac{1}{M_\mathrm{B}}\right)^2 \grad_R \cdot \grad_\text{el}\,,
\end{equation}
where the electron positions are taken relative to the geometric
center of the nuclei.  Moreover, \(\mu_\mathrm{n} = M_\mathrm{A}
M_\mathrm{B} / (M_\mathrm{A} + M_\mathrm{B})\) is the reduced nuclear
mass for the nuclei A and B; the internuclear distance is given by
\(\vb{R} = \vb{R}_\mathrm{AB} = \vb{R}_\mathrm{A} - \vb{R}_\mathrm{B}\) and
\(\grad_\text{el} = (\grad_1 + \grad_2)/2\) for the electrons \(1\)
and \(2\).

A consequent non-adiabatic treatment has been developed in the
framework of the non-adiabatic perturbation theory
(NAPT)~\cite{Czachorowski:2018xvk}. Here, the total wave function is
decomposed into an electronic and nuclear part, \(\psi_\mathrm{el}\)
and \(\chi\), respectively, while the non-adiabatic mixing effects are
encoded in a small deviation \(\delta\Psi_\text{na}\),
\begin{equation}\label{eq:NAPTwavefunc}
  \Psi(\vb{r}_1, \vb{r}_2, \vb{R}) = \psi_\mathrm{el}(\vb{r}_1,
  \vb{r}_2; \vb{R}) \chi(\vb{R}) + \delta\Psi_\text{na}(\vb{r}_1,
  \vb{r}_2, \vb{R})\,,
\end{equation}
such that \(\Braket{\delta \Psi_\text{na}}{\psi_\mathrm{el}} = 0\)
  and \(\ket{\psi_\mathrm{el}}\) solves the electronic Schr\"odinger
equation for the Hamiltonian~\eqref{eq:HeitlerLondon},
\begin{equation} \label{eq:bornOppenheimerSchroedingerEquation}
  H_\mathrm{el} \ket{\psi_\mathrm{el}} = \mathcal{E}^{(2,0)} (R)
  \ket{\psi_\mathrm{el}} \,.
\end{equation}
This yields the wave function \( \psi_\mathrm{el}(\vb{r}_1,
\vb{r}_2; \vb{R}) \) and the leading-order eigenvalue \(
\mathcal{E}^{(2,0)}(R) \), closely following the notation of
Reference~\cite{Czachorowski:2018xvk}. The Born--Oppenheimer energy \(
\mathcal{E}^{(2,0)}(R) \) serves as an effective potential in the
nuclear Schr\"odinger equation
\begin{equation}
  \left[ -\frac{\laplacian_R}{2\mu_\mathrm{n}} +
    \mathcal{E}^{(2,0)}(R) \right] \chi(\vb{R}) = E^{(2,0)}
  \chi(\vb{R})\,,
\end{equation}
which is solved by the wave function \( \chi(\vb{R}) \) and the
leading-order energy \( E^{(2,0)} \) of the full problem. Note that
\(\chi(\vb{R})\) can be factorized as \( \chi(\vb{R}) = \frac{u(R)}{R}
Y_J^m(\vu{R}) \) with a radial wave function \(u(R)\) and the
spherical harmonics \( Y_J^m(\vu{R}) \), resulting in the differential
equation of an anharmonic oscillator for the radial part. Hence, the
energy levels and $\chi(\vb{R})$ are characterized by the non-negative
integers \( J = 0, 1, \dots \) for the angular momentum, and \(v = 0,
1, \dots\) for the oscillatory part.

Finally, non-adiabatic, relativistic, and QED corrections as well as
finite nuclear-size effects are added perturbatively. For instance,
the first non-adiabatic correction reads~\cite{Czachorowski:2018xvk}
\begin{equation}
  E^{(2,1)} = \bra{\chi_{v,J}} \bra{\psi_\mathrm{el}}H_\mathrm{n}\ket{\psi_\mathrm{el}} \ket{\chi_{v,J}}\,.
\label{eq:non-adiabatic correction}
\end{equation}
This leads to an expansion in both the fine structure constant \( \alpha \) and the ratio \( m_\mathrm{e}/\mu_\mathrm{n} \),
\begin{equation}
  E(\alpha) = \alpha^2 \left( E^{(2,0)} +
  \frac{m_\mathrm{e}}{\mu_\mathrm{n}} \: E^{(2,1)} + \left(
  \frac{m_\mathrm{e}}{\mu_\mathrm{n}} \right)^2 \: E^{(2,2)} + \dots
  \right) + \alpha^4 \left( E^{(4,0)} +
  \frac{m_\mathrm{e}}{\mu_\mathrm{n}} \: E^{(4,1)} + \dots \right) +
  \dots\,,
\end{equation}
where all displayed terms and the leading corrections of \(
\order{\alpha^5, \alpha^6} \) are fully known, while the contribution
of \( \order{\alpha^7} \) is partly available~\cite{Puchalski:2018pdf,
  Puchalski:2019prl, Puchalski:2019pra, Puchalski:2016xy,
  Puchalski:2017pra, Puchalski:2018xy, Puchalski:2016ibj,
  Czachorowski:2018xvk}.

All existing corrections are implemented in the computer code
\texttt{H2spectre}~\cite{H2spectre}, which returns the energy levels
and transition energies of all hydrogen isotopologues. Moreover, the
program \texttt{H2Solv}~\cite{Pachucki:2016H2Solv} can be used to
determine the numerical solution of the electronic Schr\"odinger
equation~\eqref{eq:bornOppenheimerSchroedingerEquation}.

\subsection{Experimental Status} \label{subsec:experimentalStatus}

During the last decade, different precision spectroscopy methods have
been applied to measure fundamental vibrational modes of hydrogen
isotopologues with high accuracy.  For example, Doppler-free laser
spectroscopy uses the principle of two counterpropagating waves of the
same frequency resulting in a cancellation of the Doppler shift
effects, see Reference~\cite{BIRABEN2019671} for a review. The
application of this method allowed the measurement of several
vibrational energy levels of H$_2$, D$_2$ and HD up to a relative
precision of $\order{10^{-10}}$~\cite{Dickenson:2013, NIU201444,
  PhysRevLett.120.153002}.

By contrast, stimulated Raman spectroscopy uses two laser beams of
different frequencies with one frequency being scanned over. If the
frequency difference matches the energy of a physical transition, the
intensity of the Stokes line will be enhanced, as described in
Reference~\cite{doi:10.1002/jrs.5499} and references therein.  While
several lines have been measured, for instance in D$_2$ with a
relative precision of $\order{10^{-6}}$~\cite{doi:10.1002/jrs.5499},
an improvement is given by the CARS spectroscopy
technique~\cite{Schlosser_2017}.  Here instead, the anti-Stokes line
is coherently induced which---although suppressed---leads to a cleaner
measurement due to less background lines. Another advantage of the new
technique is the use of smaller probe volumes, which is especially
advantageous when dealing with a radioactive gas like Tritium.  This
allowed to record the world best spectrum of molecular Tritium
T\(_2\)~\cite{Schlosser_2017, Trivikram:2018} and recently of
DT~\cite{Lai2019DT}, see Table~\ref{tab:compT2} for the measurement of
T$_2$.

\begin{table}
  \caption{Example of the measurement of fundamental vibrational
    splittings in the T\(_2\) molecule for the \( Q(J) \) band, which
    is given by transitions from \( v = 1\) to \( v = 0 \) for a fixed
    rotational quantum number~\( J \)~\cite{Trivikram:2018}. The
    central theory values have been extracted from
    \texttt{H2spectre}~\cite{H2spectre}, while the theory
    uncertainties are calculated according to
    Equation~\eqref{eq:theoryUncertaintyEstimation}. All numbers are
    given in \si{\centi\meter^{-1}}.}\label{tab:compT2}
  \begin{center}
    \begin{tabular}{cS[table-format=4.4(1)]S[table-format=4.4(1)]S[table-format=+1.4]}
      \toprule
      Line & {experiment} & {theory} & {difference} \\
      \midrule
      \(Q(0)\) & 2464.5052 +- 0.0004 & 2464.5042 +- 0.0003 & 0.0010 \\
      \(Q(1)\) & 2463.3494 +- 0.0003 & 2463.3484 +- 0.0003 & 0.0010 \\
      \(Q(2)\) & 2461.0388 +- 0.0004 & 2461.0392 +- 0.0003 & -0.0004 \\
      \(Q(3)\) & 2457.5803 +- 0.0004 & 2457.5814 +- 0.0003 & -0.0011 \\
      \(Q(4)\) & 2452.9817 +- 0.0004 & 2452.9821 +- 0.0003 & -0.0004 \\
      \(Q(5)\) & 2447.2510 +- 0.0004 & 2447.2509 +- 0.0003 & 0.0001 \\
      \bottomrule
    \end{tabular}
  \end{center}
\end{table}

Note that there is a discrepancy between theory and experiment in some
lines of the T$_2$ spectrum, which might be explained by New Physics
effects.  However, the deviating sign of the difference \(E_\text{exp}
- E_\text{theo}\) in Table~\ref{tab:compT2} makes a New Physics
interpretation acting on the inter-particle potentials more
challenging.  In the context of DT, the authors of
Reference~\cite{Lai2019DT} performed a quick analysis of a pure Yukawa
potential among the nuclei,
\begin{equation}\label{eq:FFnucl}
  V(R; \tilde\alpha, \lambda) \sim \, \frac{\tilde\alpha
    \el^{-R/\lambda}}{R} \,,
\end{equation}
where \(\tilde\alpha\) and \(\lambda\) are the coupling strength and
interaction length introduced by a new force. Demanding compatibility
within one standard deviation, they derive an upper bound on the
interaction strength \(\tilde{\alpha} < 2 \times 10^{-8}
\alpha_{\mathrm{em}}\) for a distance of \(\lambda =
\SI{1}{\angstrom}\).  A more detailed analysis of New Physics effects
is done in this work.

\section{New Physics Coupled to Electrons and Nuclei} \label{sec:results}
The simplest incarnation of light New Physics is a Yukawa-type
exchange potential of a massive particle, see
Equation~\eqref{eq:potdev}.  This appears generically in many models
like the traditional scalar exchange, the leading contribution of a
vector mediator, or in the deconstruction of an extra dimension where
the particle ``mass'' is replaced by the inverse size of the
extra dimension, see Reference~\cite{Salumbides:2015qwa}.  While the
latter is supposed to give a universal contribution to all massive
objects rather like a \emph{fifth force} extending gravity, scalar and
vector mediators may couple with different charges to electrons and
nuclei (or quarks). Another interesting option is given by the
exchange of two fermions between the two nuclei of a molecule, such as
the two-neutrino exchange~\cite{Feinberg:1968zz}. In all these cases,
one might expect a measurable effect in molecular spectra if the new
particle's mass is of $\order{\si{\kilo\electronvolt}}$.

There are very strong but indirect constraints on these types of new
interactions. Star cooling gives an efficient constraint from both the
sun and red giants, excluding a large part of the parameter space in
the \si{\kilo\electronvolt}-regime~\cite{Raffelt:1990yz,
  Raffelt:1996wa, Raffelt:2006cw}.  However, it is not clear up to
which masses these bounds are valid.  It is nevertheless interesting
to study this mass range with direct laboratory experiments since they
are exclusively accessible in molecular and atomic spectroscopy. Thus,
they directly probe new nucleus-nucleus, electron-nucleus and
electron-electron interactions.

We assume a generic New Physics potential being present among all
particles in the molecule, that is between all combinations of the two
nuclei A and B with masses \(m_{\mathrm{A},\mathrm{B}}\) and two
electrons \(1\) and \(2\) with mass \(m_\mathrm{e}\).  For instance,
adding a new Yukawa interaction of a mediator of mass $m$ to the
Coulomb force, the full potential in the notation of
Equation~\eqref{eq:HeitlerLondon} reads
\begin{equation}\label{eq:fullpot}
  \begin{alignedat}{6}
  &V_\text{NP-full} (r_1, r_2, r_\mathrm{A}, r_\mathrm{B}) = \alpha_\mathrm{em} &&\:\bigg\{&&
  \left(-1+\frac{g_\text{eN}}{4\pi\alpha_\mathrm{em}} \el^{-mr_\mathrm{1A}}\right)
  \frac{1}{r_\mathrm{1A}} &&\:{}+{}\:&&
  \left(-1+\frac{g_\text{eN}}{4\pi\alpha_\mathrm{em}} \el^{-mr_\mathrm{2B}}\right)  \frac{1}{r_\mathrm{2B}}\\
  &&&\:{}+{}\:&& \left(-1+\frac{g_\text{eN}}{4\pi\alpha_\mathrm{em}} \el^{-mr_\mathrm{1B}}\right)
  \frac{1}{r_\mathrm{1B}} &&\:{}+{}\:&&
  \left(-1+\frac{g_\text{eN}}{4\pi\alpha_\mathrm{em}} \el^{-mr_\mathrm{2A}}\right)
  \frac{1}{r_\mathrm{2A}} \\
  &&&\:{}+{}\:&& \left(1+ \frac{g_\text{ee}}{4\pi\alpha_\mathrm{em}} \el^{-mr_{12}} \right)
  \frac{1}{r_{12}} &&\:{}+{}\:&&
  \left(1+ \frac{g_\text{NN}}{4\pi\alpha_\mathrm{em}} \el^{-mR_\mathrm{AB}} \right)
  \frac{1}{R_\mathrm{AB}}
  \: \bigg\}\,
\end{alignedat}
\end{equation}
with a New Physics coupling between electrons and nuclei
\(g_\text{eN}\), electrons and electrons \(g_\text{ee}\), and nuclei
and nuclei \(g_\text{NN}\). In principle, each \(g_{ij}\) can have
both signs and thus work either attractive or repulsive irrespective
of the electric charge. However, \(g_\text{NN}\) and \(g_\text{ee}\)
should be positive bearing in mind that all coupling strengths
$g_{ij}$ are rather multiplicative couplings if expressed in terms of
fundamental interactions $y_i$ as $g_{ij} \sim y_i y^*_j$.

The \emph{fifth force} analyses of
References~\cite{Salumbides:2013aga} and~\cite{Lai2019DT} only
constrain the last term of Equation~\eqref{eq:fullpot} with the
replacement \(\frac{g_\text{NN}}{4\pi} \to \alpha_5\).  In the
following, we will extent this analysis by discussing also the other
terms in Equation~\eqref{eq:fullpot} and more types of potentials.

\subsection{Implementation of the New Physics Corrections}

In order to estimate the full impact of New Physics on the spectra, we
set all but one coupling $g_{ij}$ to zero.  For a given New Physics
potential $V_\mathrm{NP}$, the energy correction \( \Deltaup
E^\mathrm{NP}_{v,J}\) of a rovibrational level \( (v,J) \) is
calculated in first-order perturbation theory by evaluating the matrix
element
\begin{equation}
  \Deltaup E^\mathrm{NP}_{v,J} = \bra{\chi_{v,J}} \bra{\psi_\mathrm{el}} V_\mathrm{NP} \ket{\psi_\mathrm{el}} \ket{\chi_{v,J}} \,,
\end{equation}
so that the full energy reads
\begin{equation}
  E^\mathrm{NP}_{v,J} = E^\mathrm{SM}_{v,J} + \Deltaup E^\mathrm{NP}_{v,J} \,.
\end{equation}
Here, \( E^\mathrm{SM}_{v,J} \) describes the Standard Model
prediction which, including its theoretical uncertainty \( \delta
E^\mathrm{SM}_{v,J} \), can be extracted from the computer code
\texttt{H2spectre}~\cite{H2spectre}. For the evaluation of the New
Physics shift \( \Deltaup E^\mathrm{NP}_{v,J} \), we use the same
unperturbed states \( \ket{\psi_\mathrm{el}}\ket{\chi_{v,J}} \) that
also enter the computation of all corrections in the Standard Model
calculation, see Equation~\eqref{eq:non-adiabatic correction}.

The case of a pure nuclear force is straightforward. Here, the
electronic part \( \ket{\psi_\mathrm{el}} \) of the wave function
evaluates to $1$, leaving
\begin{equation}
  \Deltaup E^\mathrm{NP}_{v,J} = \bra{\chi_{v,J}} V_\mathrm{NP} \ket{\chi_{v,J}} \,.
\end{equation}
We extract the nuclear wave function \( \chi_{v,J}(R) \) from
\texttt{H2spectre} in a discrete value representation (DVR) with grid
spacing \( \Deltaup R \). Analogously to the \texttt{H2spectre}
computation of the higher-order Standard Model
corrections~\cite{H2spectre}, we use
\begin{equation}
  \label{eq:dvrEnergyCorrection}
  \Deltaup E^\mathrm{NP}_{v,J} = \Deltaup R \cdot \sum_i V_{\mathrm{NP}}^i \cdot \left( \chi_{v,J}^i \right)^2 \,,
\end{equation}
with \( \chi_{v,J}^i = \chi_{v,J}(R_i) \) and \( V_{\mathrm{NP}}^i =
V(R_i) \) being the nuclear wave function and potential evaluated at
the DVR grid points \( R_i \), respectively.

In case of a force that also couples to electrons, the electronic
matrix element
\begin{equation}
  \mathcal{E}^\mathrm{el}_\mathrm{NP}(R) = \bra{\psi_\mathrm{el}} V_\mathrm{NP} \ket{\psi_\mathrm{el}}
\end{equation}
needs to be evaluated first since the electronic wave function depends
on the nuclear separation \( R \).  We extract this wave function in
the symmetric James--Coolidge basis, as specified in
Equation~\eqref{eq:James-Coolidge basis}, from the publicly available
code \texttt{H2SOLV}~\cite{Pachucki:2016H2Solv}.  In particular, we
fix the nuclear distance \( R \) and minimize the energy expectation
value of the wave function as computed by \texttt{H2SOLV} with respect
to the variational parameter \( u \) defined in
Reference~\cite{Pachucki:2016H2Solv}.  Using the coefficients of the
basis expansion for the minimal energy expectation value, we compute
the electronic matrix element by numerical integration.  To avoid the
time-consuming numerical integration at each parameter point \( (R, m,
g) \), we evaluate the electronic matrix element on a grid in \( (R,
m) \) only, since the coupling \( g \) factorizes in each case.  The
full dependence of \( \mathcal{E}^\mathrm{el}_\mathrm{NP}(R) \) on \(
R \) and \( m \) is afterwards reconstructed by interpolation with
splines of degree two.  \(\mathcal{E}^\mathrm{el}_\mathrm{NP}(R)\)
obtained in this way serves as an effective potential for the nuclei
in the same manner as the relativistic corrections do in the
\texttt{H2spectre} computation. Consequently, the New Physics
contribution \( \Deltaup E_\mathrm{NP} \) is calculated using
Equation~\eqref{eq:dvrEnergyCorrection} with \(V_\mathrm{NP}(R)\)
replaced by \(\mathcal{E}^\mathrm{el}_\mathrm{NP}(R)\).

There is an additional complication for spin-dependent potentials when
coupled to nuclei. In order to comply with the Pauli principle, the
nuclear spin state \( \ket{f_1, m_{\mathrm{f},1}, f_2,
  m_{\mathrm{f},2}} \) depends on the angular momentum quantum number
\( J \). Since the leading-order energy is independent of the magic
quantum number \( m_{\mathrm{f},i} \) of the nucleus \( i \), this
leads to a degeneracy and, hence, we need to use degenerate
perturbation theory to calculate the energy correction. In this case,
the New Physics energy shift \( \Deltaup E^\mathrm{NP}_{v,J} \) for
the ground state is determined by the minimal eigenvalue of the
perturbation in the degenerate subspace.

Finally, the energy \( \Deltaup E^\mathrm{NP}_{(v_1,J_1) \rightarrow
  (v_2,J_2)} \) for the transition from a level \( (v_1, J_1) \) to
the level \( (v_2, J_2) \) is given by
\begin{equation}
  \Deltaup E^\mathrm{NP}_{(v_1,J_1) \rightarrow (v_2,J_2)} = E^\mathrm{NP}_{v_2,J_2} - E^\mathrm{NP}_{v_1,J_1} \,.
\end{equation}
We expect the size of the New Physics contribution to be of order of
the uncertainty \( \delta E^\mathrm{SM}_{v,J} \) of the Standard Model
calculation. Since the theoretical error in the New Physics energy
shift should be much smaller than the contribution itself, \( \delta
E^\mathrm{NP}_{v,J} \ll \Deltaup E^\mathrm{NP}_{v_2,J_2} \sim \delta
E^\mathrm{SM}_{v,J} \), we approximate the overall uncertainty of the
level energy to be \( \delta E^\mathrm{SM}_{v,J} \), that is
\begin{equation}
  \delta E^\mathrm{NP}_{v,J} = \delta E^\mathrm{SM}_{v,J} + \delta E^\mathrm{NP}_{v,J} \approx \delta E^\mathrm{SM}_{v,J} \,.
\end{equation}
In contrast to the error estimate for transitions in
\texttt{H2spectre}, we linearly add the uncertainties of the two
corresponding levels to get a more conservative theoretical
uncertainty \( \delta \Deltaup E^\mathrm{NP}_{(v_1,J_1) \rightarrow
  (v_2,J_2)} \) for the transition energy,
\begin{equation} \label{eq:theoryUncertaintyEstimation}
  \delta \Deltaup E^\mathrm{NP}_{(v_1,J_1) \rightarrow (v_2,J_2)} = \delta E^\mathrm{SM}_{v_1,J_1} + \delta E^\mathrm{SM}_{v_2,J_2} \,.
\end{equation}
Given an experimental measurement $\Deltaup
E^{\mathrm{exp}}_{(v_1,J_1) \rightarrow (v_2,J_2)}$ for a transition
$(v_1,J_1) \rightarrow (v_2,J_2)$ with an uncertainty
$\sigma^\mathrm{exp}_{(v_1,J_1) \rightarrow (v_2,J_2)}$, we require
the theoretical prediction including New Physics effects to lie within
the interval
\begin{equation}
\label{eq:criterion for limits}
    [\Deltaup E^{\mathrm{exp}} - 3\,\sigma^\mathrm{exp} - \delta \Deltaup E^\mathrm{NP},
    \Deltaup E^{\mathrm{exp}} + 3\,\sigma^\mathrm{exp} + \delta \Deltaup E^\mathrm{NP}]\,,
\end{equation}
for each transition, suppressing the indices for
clarity.  For a given molecule and mass of the new mediator, this
criterion allows to derive upper bounds on the couplings $g_{ij}$ by
combining all measurements listed in Appendix~\ref{app:Experimental
  data}.

\subsection{Scalar and Pseudoscalar Potentials}
Many New Physics scenarios comprise light scalar or pseudoscalar
fields, for instance an additional light Higgs
boson~\cite{Silveira:1985rk, McDonald:1993ex} or as remnants of
(softly or spontaneously) broken continuous global symmetries and thus
(Pseudo-)Nambu--Goldstone bosons, like the axion~\cite{Peccei:1977hh,
  Peccei:1977ur, Wilczek:1977pj, Weinberg:1977ma, Preskill:1982cy}
or the Majoron~\cite{Chikashige:1980ui}.

Complete spin-dependent potentials for various mediator particles have
been summarized in~\cite{Fadeev:2018rfl}. In particular, potentials
for massive scalar (S) or pseudoscalar (P) mediators \(\phi\) with
mass $m$ between two fermions a and b with masses $m_\mathrm{a,b}$,
are given by the expressions
\begin{subequations}\label{eq:scalarpots}
  \begin{align}
    V_{\mathrm{S}}(\vb{r}) =\;& - g^\mathrm{S}_{\mathrm{ab}}
                      \frac{\el^{-m r}}{4\pi r} \,, \\
    V_{\mathrm{P}}(\vb{r}) =\;& - g^\mathrm{P}_{\mathrm{ab}}
                      \frac{m^2}{4m_{\mathrm{a}} m_{\mathrm{b}}} \bigg[
                      \left( \bm{\sigma}_\mathrm{a} \cdot \bm{\sigma}_\mathrm{b} \right)
                      \left(\frac{1}{m^2 r^2} + \frac{1}{m r} +
                      \frac{4\pi r}{3 m^2} \delta^{(3)}(\vb{r}) \right) \nonumber \\
    & \qquad - \left(
                      \bm{\sigma}_\mathrm{a} \cdot \vu{r} \right) \left(
                      \bm{\sigma}_\mathrm{b} \cdot \vu{r} \right) \left( 1 +
                      \frac{3}{m^2 r^2} + \frac{3}{m r} \right)
                      \bigg] \frac{\el^{-m r}}{4\pi r}\,,
  \end{align}
\end{subequations}
with the (pseudo)scalar couplings
\(g_{\mathrm{a}\mathrm{b}}^{\mathrm{S(P)}}\) and the spin Pauli matrices
\(\bm{\sigma}_\mathrm{a,b} \).

Note that the pseudoscalar interaction is suppressed by the masses of
the interacting particles. For this reason, we do not expect strong
limits from molecular systems for pseudoscalar interactions. Moreover,
these potentials have been derived between spin-\(\frac{1}{2}\)
fermions. Nevertheless, the spin-independent scalar potential can also
be applied to a force between spin-1 bosons like the deuteron, while
the pseudoscalar is to be used for spin-\(\frac{1}{2}\) particles
only.

\begin{figure}[p!]
  \vspace{-50pt}
  \captionsetup[subfigure]{justification=centering}

  \begin{subfigure}{.45\textwidth}
    \includegraphics[width=\textwidth]{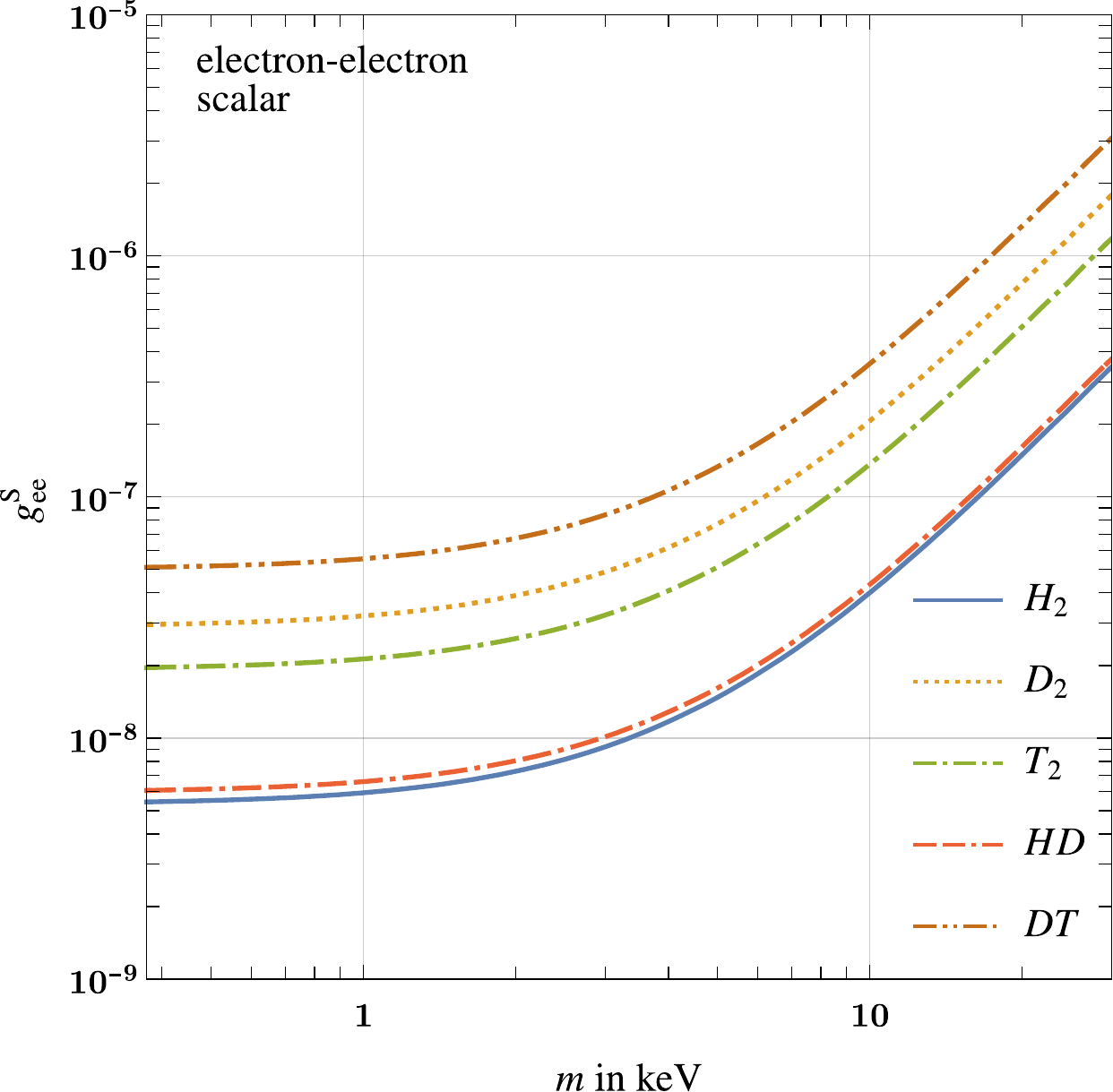}
    \subcaption{Scalar electron--electron interaction.}
    \label{fig:ee-scalar}
  \end{subfigure} \hfill
  \begin{subfigure}{.45\textwidth}
    \includegraphics[width=\textwidth]{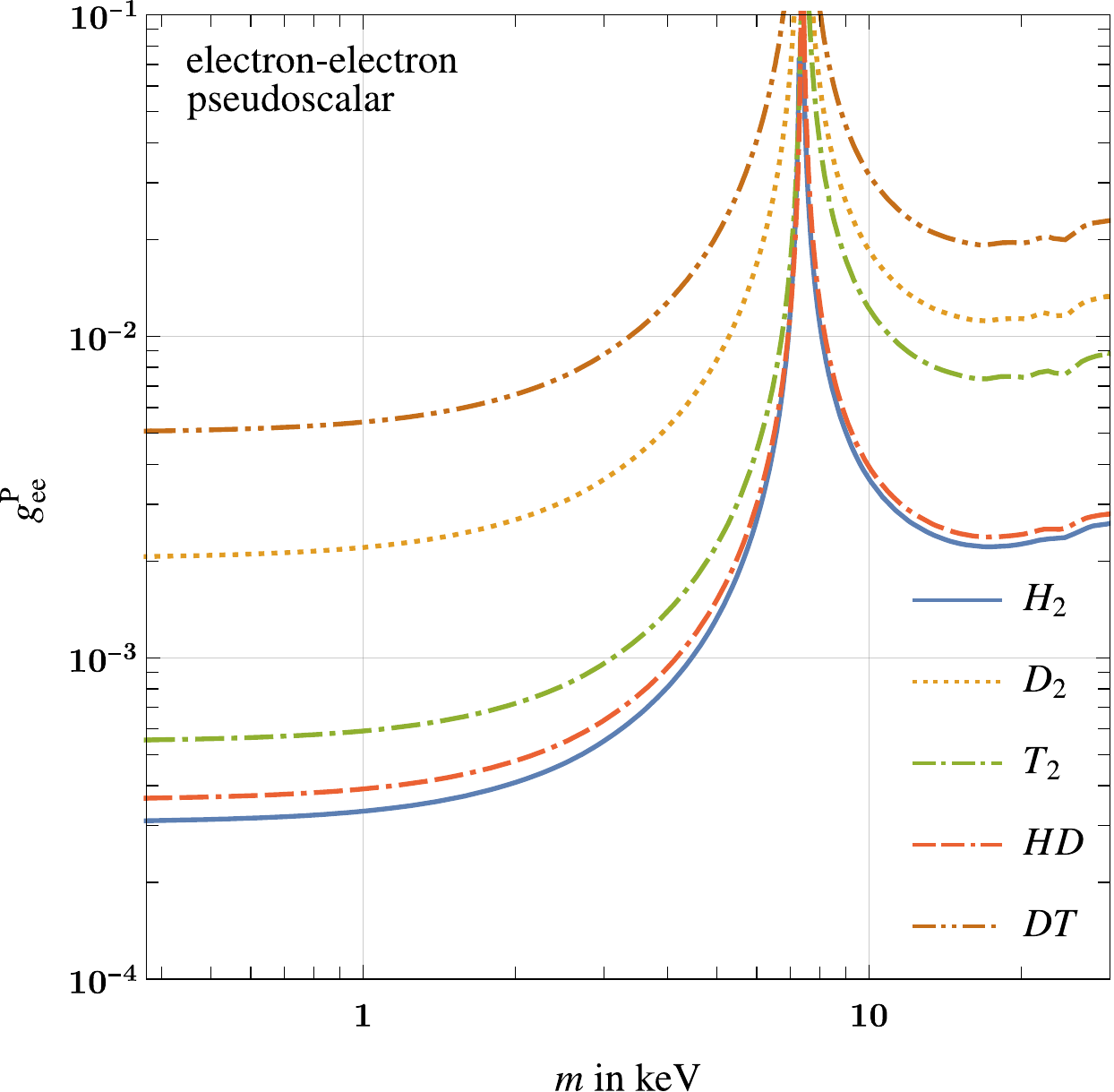}
    \subcaption{Pseudoscalar electron--electron interaction.}
    \label{fig:ee-pseudoscalar}
  \end{subfigure}
  \par\medskip
  \begin{subfigure}{.45\textwidth}
    \includegraphics[width=\textwidth]{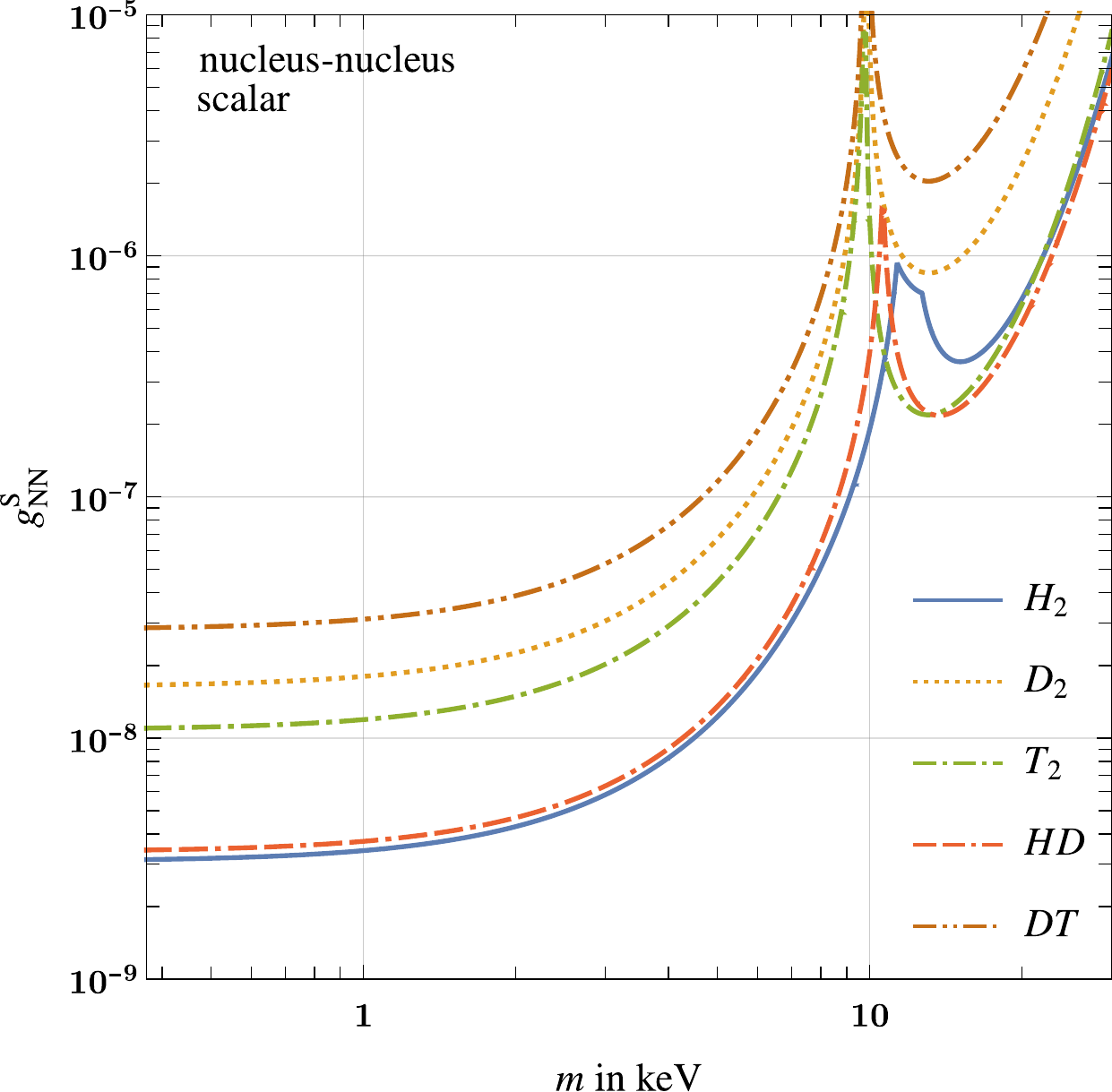}
    \subcaption{Scalar nucleus--nucleus interaction.\\\ }
    \label{fig:nn-scalar/vector}
  \end{subfigure} \hfill
  \begin{subfigure}{.45\textwidth}
    \includegraphics[width=\textwidth]{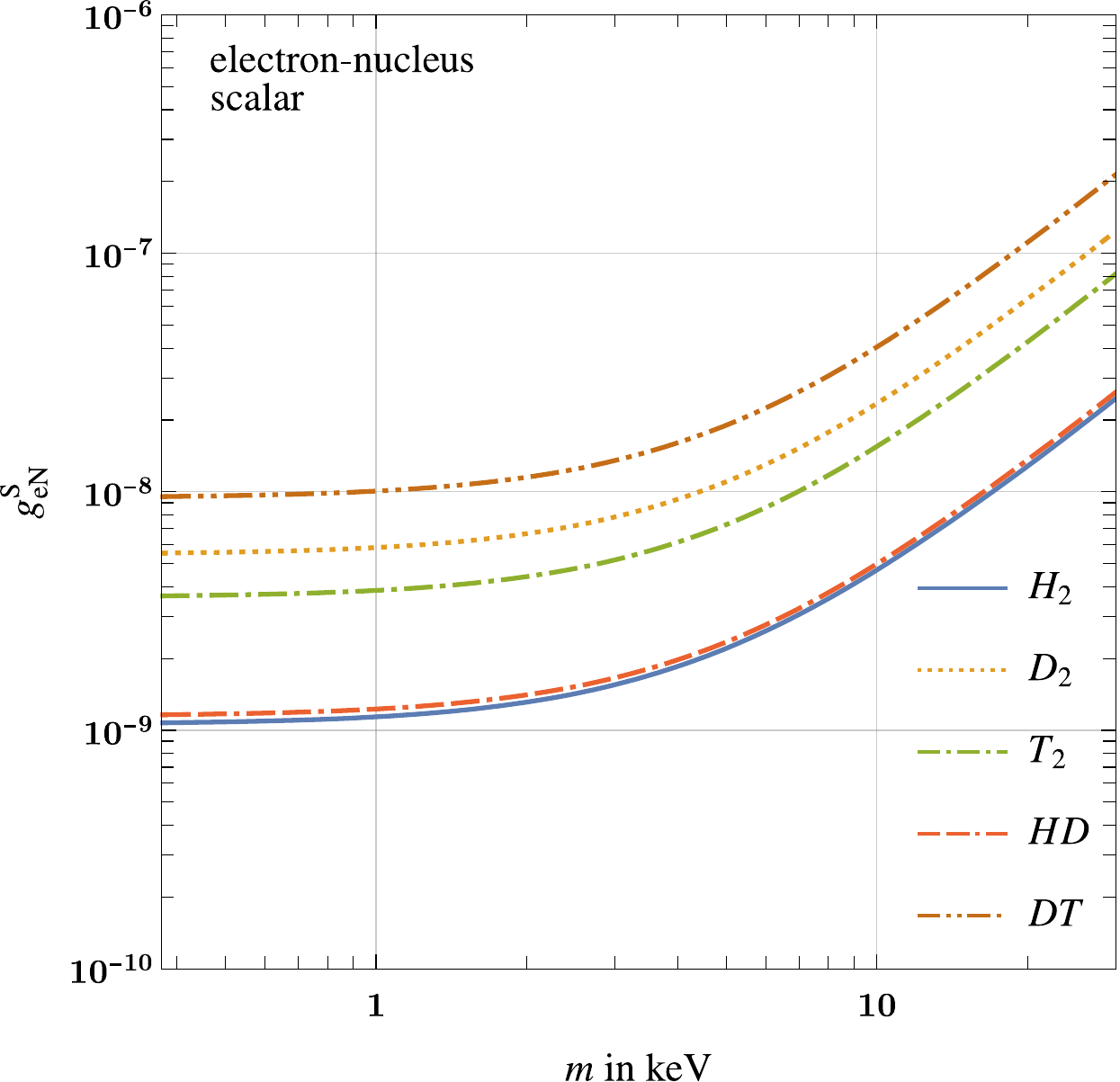}
    \subcaption{Scalar electron--nucleus interaction with positive
      coupling \( g^\mathrm{S}_\mathrm{eN} > 0 \).}
    \label{fig:en-scalar-positive}
  \end{subfigure}
  \par\medskip
  \begin{subfigure}{\linewidth}
    \centering
    \includegraphics[width=0.405\textwidth]{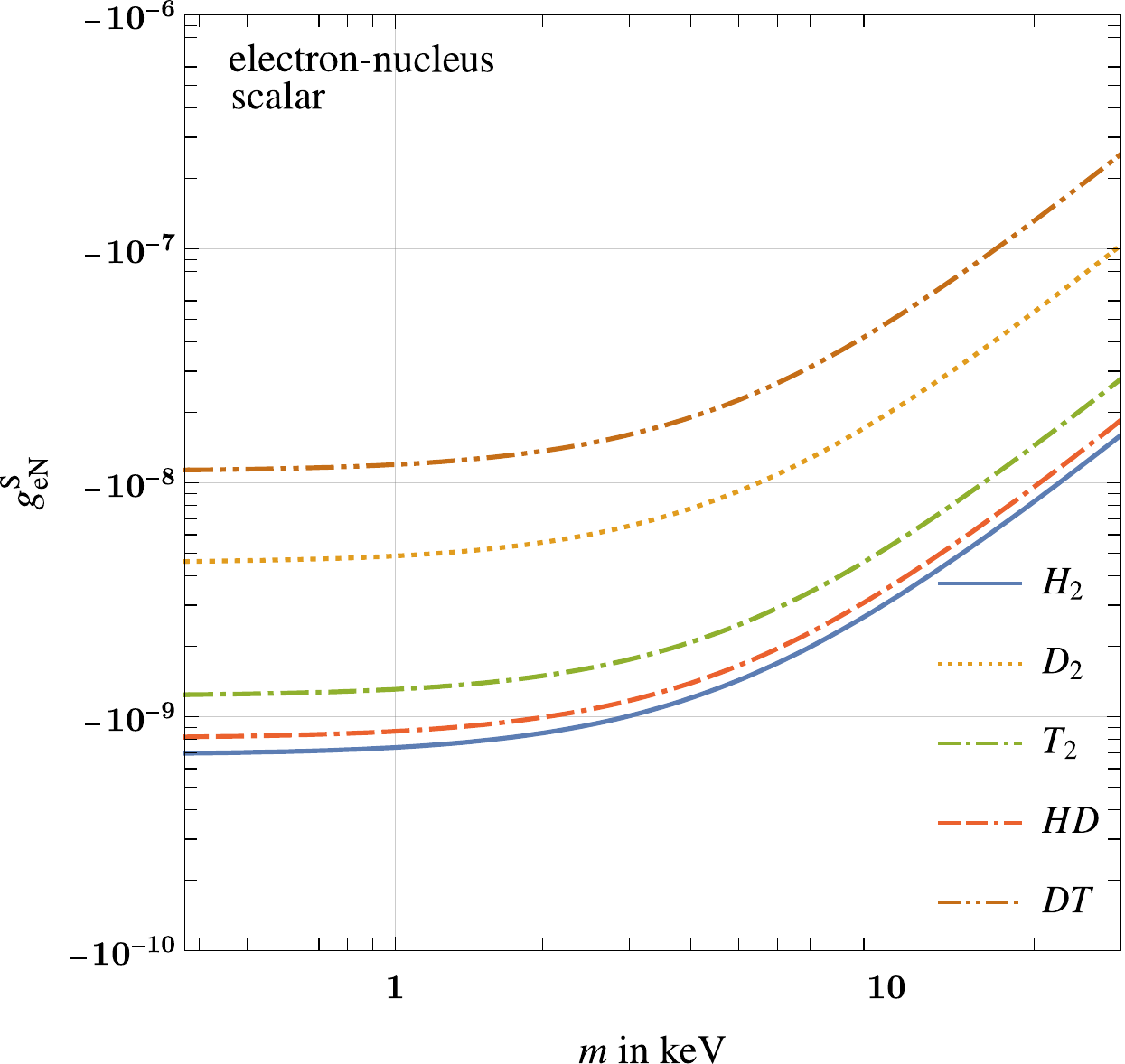}
    \subcaption{Scalar electron--nucleus interaction with negative
      coupling \( g^\mathrm{S}_\mathrm{eN} <0 \).}
    \label{fig:en-scalar-negative}
  \end{subfigure}

  \caption{Exclusion limits in mass--coupling plane on forces mediated
    by scalar or pseudoscalar particles. For each molecule, the
    corresponding upper limit results from the combination of all
    available measurements. The area above the curves is excluded.}
  \label{fig:scalar-pseudoscalar}
  \par\medskip
  \hrule
\end{figure}

Applying the criterion in Equation~\eqref{eq:criterion for limits}, we
derive upper limits on the couplings $g^{\mathrm{S, P}}_{\mathrm{ab}}$
shown in the mass--coupling plane, see
Figure~\ref{fig:scalar-pseudoscalar}.  For a scalar interaction
between electrons, cf. Equation~\eqref{eq:scalarpots}, H$_2$ and HD
molecules constrain the scalar coupling $g^\mathrm{S}_\mathrm{ee}$ up
to $\order{10^{-8}}$, see Figure~\ref{fig:ee-scalar}.  By contrast,
the coupling of a pseudoscalar mediator is weakly constrained,
$g^\mathrm{P}_{\mathrm{ee}}\sim\order{10^{-3}}$ for $m\sim
\SI{1}{keV}$, meeting the expectation of a suppression by
$m^2/m_\mathrm{e}^2\sim 10^{-6}$ relative to the scalar case, as shown in
Figure~\ref{fig:ee-pseudoscalar}. It can be seen that the bounds for
the pseudoscalar coupling become ineffective at about
\SI{7}{keV}. This happens when the New Physics contribution approaches
zero and eventually changes its sign as a consequence of an internal
cancellation between the terms with different spin structure. This is
an interesting feature which might be resolved using polarized probes.

For a pure nucleus--nucleus interaction, a pseudoscalar contribution
is even more suppressed by $m^2/m_\mathrm{N}^2 \sim 10^{-11}$, thus,
we only consider the scalar potential.  The corresponding limits in
the mass-coupling plane are shown in
Figure~\ref{fig:nn-scalar/vector}. Kinks in the plot are an artefact
of the combination of several measured lines indicating that another
measurement becomes more efficient in the exclusion of parameter
space. While the dominant limits of $\order{10^{-8}}$ for masses
$\lesssim \SI{10}{\kilo\electronvolt}$ arise from H$_2$ and HD
transitions, T$_2$ measurements are most constraining for larger
masses.  Note that our limits are weaker than the ones presented in
References~\cite{Salumbides:2013aga,Lai2019DT} as a consequence of a
more conservative exclusion criterion by allowing for \(3\sigma\)
deviations.

In the case of an electron--nucleus interaction the spin matrix
elements for the electronic ground state vanish and thus only the
  scalar interaction survives.  Since the relative sign of the
electron and nucleus coupling $g^\mathrm{S}_\mathrm{eN}$ is not fixed,
we show the exclusion limits for both signs in
Figures~\ref{fig:en-scalar-positive} and~\ref{fig:en-scalar-negative}.
As in the electron--electron case, the strongest constraints are again
given by the transitions measured in H$_2$ and HD molecules with upper
limits on the coupling $\abs{g^\mathrm{S}_\mathrm{eN}}$ up to
$\order{10^{-8}}$ and $\order{10^{-9}}$ for a positive and
negative coupling, respectively.  Compared to the electron--electron
and nucleus--nucleus case, the slightly better constraints are expected
because of the four possible combinations of electrons and nuclei.
Note that there should be another enhancement due to
$g^\mathrm{S}_\mathrm{eN}$ also implying a $g^\mathrm{S}_\mathrm{ee}$
and $g^\mathrm{S}_\mathrm{NN}$ coupling, however, the order of
magnitude will not change.

\subsection{Vector and Axialvector Exchange Potentials}
There are different options of introducing a new (axial)vector
coupling. One possibility is via kinetic mixing with a ``dark''
photon, where a new ``dark'' \(U(1)\) gauge field described by the
field-strength tensor \(F'_{\mu\nu} = \partial_\mu A'_\nu -
\partial_\nu A'_\mu\) mixes with the electromagnetic photon through a
Lagrangian of the type~\cite{Holdom:1985ag}
\begin{equation}\label{eq:kinetmix}
  \Lag_\text{kin-mix} = -\frac{1}{4} F^{\mu\nu} F_{\mu\nu} - \frac{1}{4}
  F^{\prime\mu\nu} F'_{\mu\nu} - \frac{1}{2} F^{\mu\nu} F'_{\mu\nu}
  \,.
\end{equation}
Another possibility involves the Stueckelberg mechanism where
additionally a light axion-like field is
present~\cite{Stueckelberg:1900zz, Kors:2004dx}. In this case, the
heavy vector is usually referred to a \(Z'\) boson and thus supposed
to have a mass in the \si{\giga\electronvolt} regime rather than
\si{\kilo\electronvolt}.

The presence of a light spin-1 mediator with vector and axialvector
couplings $g^\mathrm{V,A}_\mathrm{ab} =
g^\mathrm{V,A}_{\mathrm{a}}g^\mathrm{V,A}_{\mathrm{b}}$ of the type
\(A_\mu^\prime \; \bar \psi \; \gamma^\mu \left( g^\mathrm{V}_{\psi} +
\im \gamma_5 \; g^\mathrm{A}_{\psi} \right) \psi\) and a mass \(m\)
leads to non-relativistic potentials~\cite{Fadeev:2018rfl}
\begin{subequations} \label{eq:vectorpots}
  \begin{align}
    V_\mathrm{V} (\vb{r}) =\;&  \frac{g^\mathrm{V}_\mathrm{ab}}{4\pi} \;
                          \frac{\el^{-m r}}{r}
               \bigg \lbrace 1 + \nonumber\\
    & \qquad \frac{m^2}{4 m_\mathrm{a} m_{\mathrm{b}}} \bigg[ \bm{\sigma}_\mathrm{a}
     \cdot \bm{\sigma}_\mathrm{b} \left(\frac{1}{m^2 r^2} + \frac{1}{m r}
      + 1 - \frac{8\pi r}{3m^2} \delta(\vb{r}) \right) \nonumber\\
    &\qquad\qquad\qquad\qquad - (\bm{\sigma}_\mathrm{a} \cdot \vu{r})(\bm{\sigma}_\mathrm{b} \cdot \vu{r}) \left(
    \frac{3}{m^2 r^2} + \frac{3}{m r} + 1 \right)
      \bigg] \bigg \rbrace \,, \label{eq:potveve}\\
    V_\mathrm{A} (\vb{r}) =\;& - \frac{g^\mathrm{A}_\mathrm{ab}}{4\pi}  \frac{\el^{-mr}}{r} \bigg \lbrace
                          \; \bm{\sigma}_\mathrm{a} \cdot
                          \bm{\sigma}_\mathrm{b} \bigg[ 1 +
                          \frac{1}{m^2 r^2} +
                          \frac{1}{m r} + \frac{4\pi r}{3
                          m^2} \delta^{(3)}(\vb{r}) \bigg]
                           \nonumber \\
                           &\qquad\qquad\qquad\qquad
                          -
                          (\bm{\sigma}_\mathrm{a} \cdot \vu{r})
                          (\bm{\sigma}_\mathrm{b} \cdot \vu{r}) \bigg[ 1+
                          \frac{3}{m^2 r^2} +
                          \frac{3}{m r} \bigg] \bigg \rbrace
                          \,. \label{eq:potaxax}
  \end{align}
\end{subequations}
Here, $\bm{\sigma}_i$ and \(\vu{r}\) are the Pauli matrices of particle
$i$ and the unit vector pointing in the direction
between the two fermions a and b, respectively.

The dominant spin-independent effect can be found from the
\(V_\mathrm{V}\) potential above, being exactly the Yukawa-type
potential mentioned earlier. Note that the spin-dependent vector
interactions are suppressed by the fermion masses.  Thus, the leading
contribution for a vector mediator is given by the Yukawa potential
\begin{equation}\label{eq:leadvec}
  V_\text{V}(\vb{r}) = \pm \frac{g^\text{V}_{\mathrm{ab}}
    }{4\pi} \frac{\el^{-mr}}{r}\,.
\end{equation}
In contrast to the pseudoscalar case, the axial vector interaction is
not suppressed by the inverse fermion masses. The contribution from
the axialvector potential is expected to be of the same size as the
leading vector contribution, where additionally the dependence on the
mediator mass plays a more significant role. As for the pseudoscalar
case, the potentials have been derived for an interaction among
fermions and, hence, we only consider the leading Yukawa-type
contribution for bosonic nuclei.

\begin{figure}[t!]
  \captionsetup[subfigure]{justification=centering}

  \begin{subfigure}{.45\textwidth}
    \includegraphics[width=\textwidth]{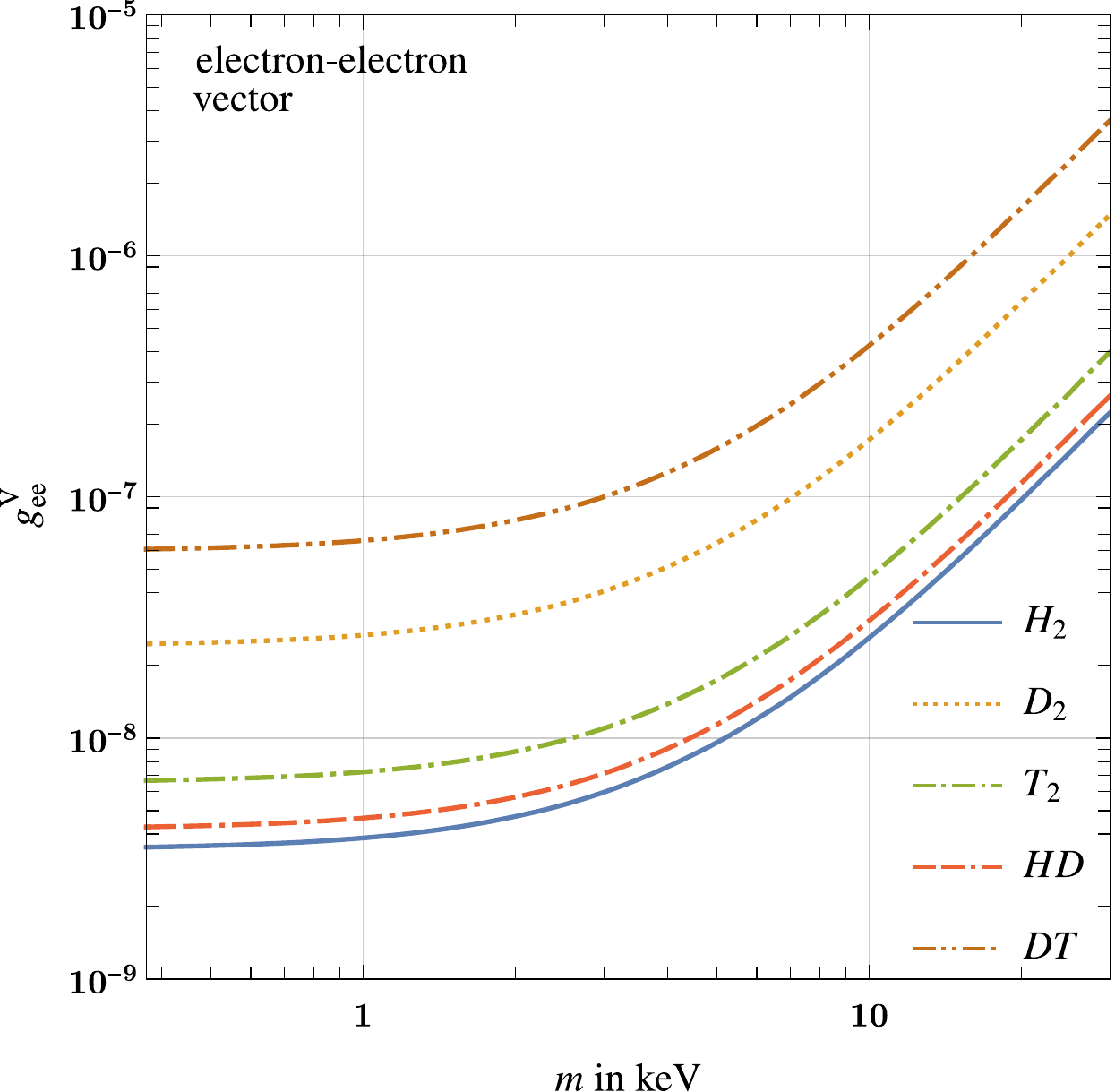}
    \subcaption{Vector electron--electron interaction.}
    \label{fig:ee-vector}
  \end{subfigure} \hfill
  \begin{subfigure}{.45\textwidth}
    \includegraphics[width=\textwidth]{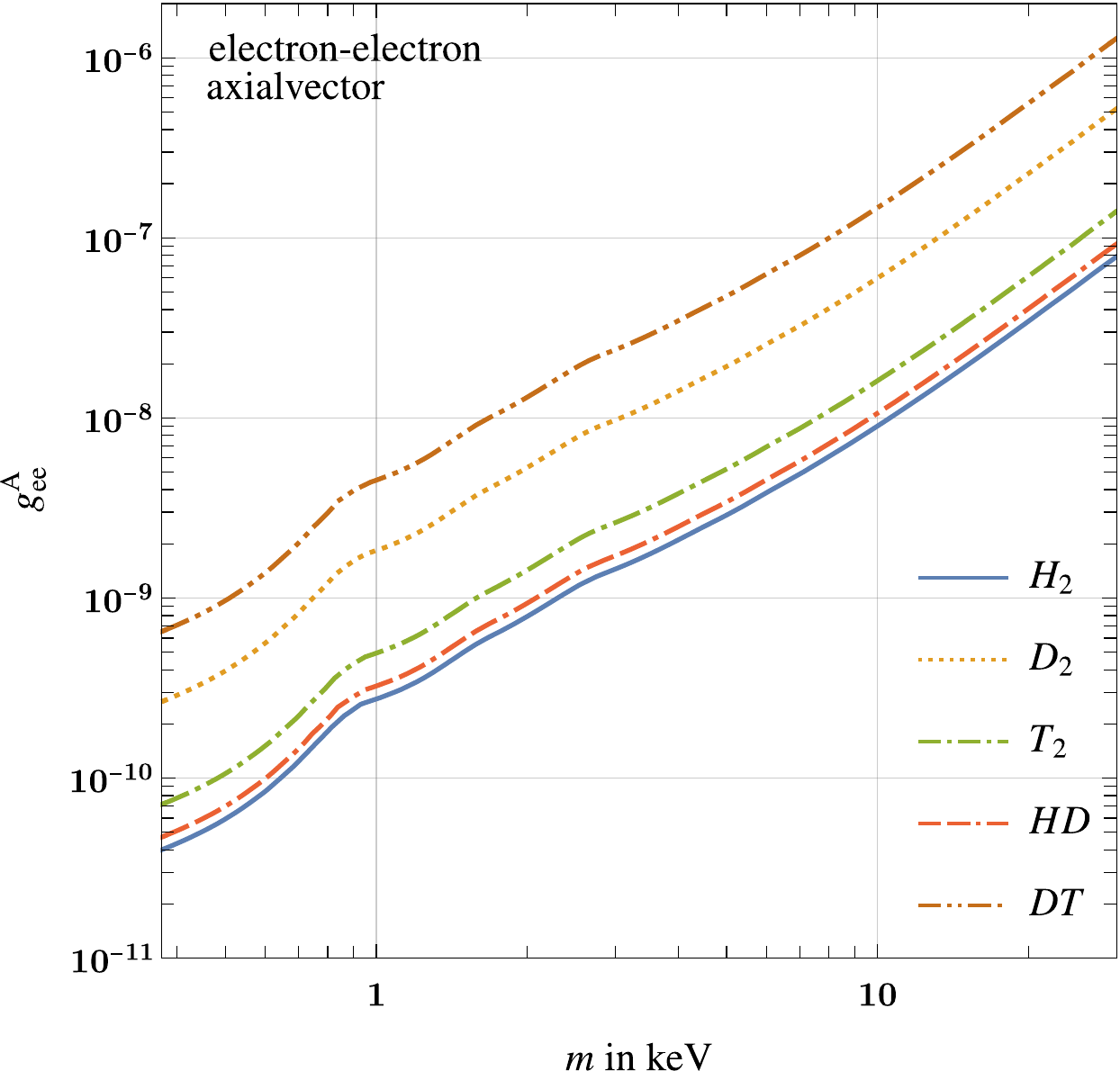}
    \subcaption{Axialvector electron--electron interaction.}
    \label{fig:ee-axialvector}
  \end{subfigure}
  \par\medskip
  \begin{subfigure}{\linewidth}
    \centering
    \includegraphics[width=.45\textwidth]{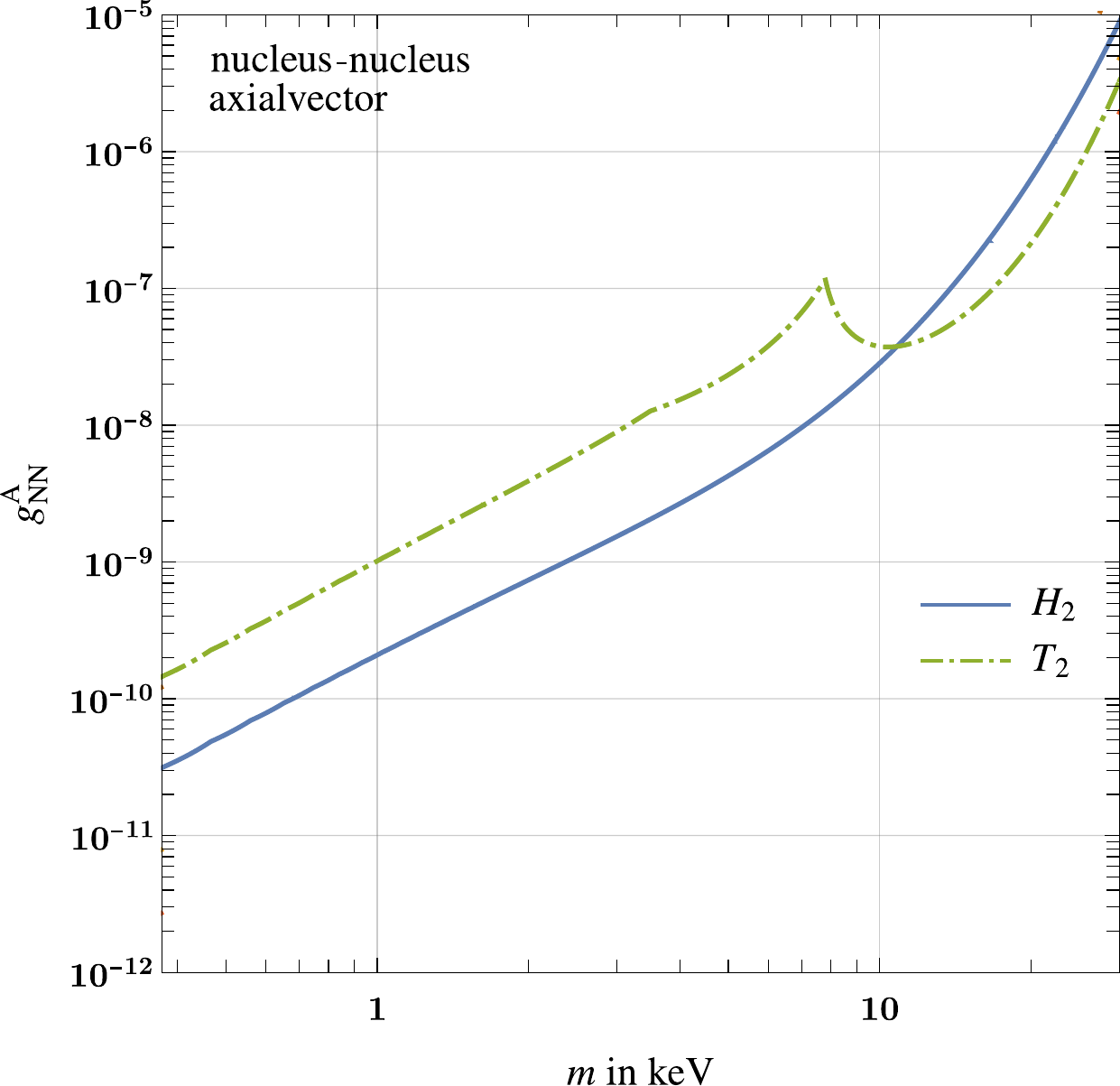}
    \subcaption{Axialvector nucleus--nucleus interaction.}
    \label{fig:nn-axialvector}
  \end{subfigure}
  \caption{Exclusion limits in mass--coupling plane on forces mediated
    by vector or axialvector particles. For each molecule, the
    corresponding upper limit on the coupling constants results from
    the combination of all available measurements.}
  \label{fig:vector interactions}
  \par\medskip
  \hrule
\end{figure}

Like in the scalar interaction, the strongest constraints for a pure
electronic force are given by the measurement of H$_2$ and HD
transition, see Figures~\ref{fig:ee-vector}
and~\ref{fig:ee-axialvector}. We find an upper limit on the coupling
$g^\mathrm{V,A}_\mathrm{ee}$ of $\order{10^{-8}}$ and
$\order{10^{-10}}$ for the vector and axialvector potential,
respectively, for masses around \SI{1}{\kilo\electronvolt}.

Regarding the nucleus--nucleus force, the additional terms in the
vector contribution are suppressed by two powers of the nuclear mass
and, therefore, the limits coincide with the ones for the scalar
potential shown in Figure~\ref{fig:nn-scalar/vector}.  Bounds from the
axialvector exchange are again stronger by two orders of magnitude
yielding an upper bound on the coupling $g_\mathrm{NN}^\mathrm{A}$ of
$\order{10^{-10}}$, see Figure~\ref{fig:nn-axialvector}. Since we do
not consider the bosonic nuclei D\(_2\) and HD, the best limits are
now given by H\(_2\) measurements for masses below \( \SI{10}{keV} \)
and by T\(_2\) lines for larger masses.

Analogously to the pseudoscalar electron--nucleus interaction, the
spin-dependent terms vanish for the electronic ground state of H$_2$
isotopologues. As a consequence, the bounds on the vector potential
are the same as for the scalar case, see
Figures~\ref{fig:en-scalar-positive} and~\ref{fig:en-scalar-negative},
while an axialvector force vanishes entirely.

\subsection{Singular Potentials: Effective Contact Interactions}

\begin{figure}[tb]
  \begin{minipage}{.4\textwidth}
    \includegraphics[width=\textwidth]{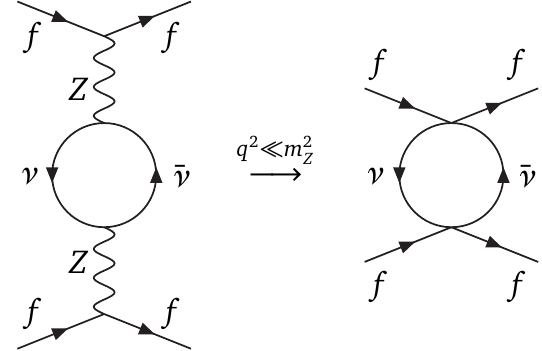}
  \end{minipage}%
  \hfill\vline\hfill
  \begin{minipage}{.4\textwidth}
    \includegraphics[width=\textwidth]{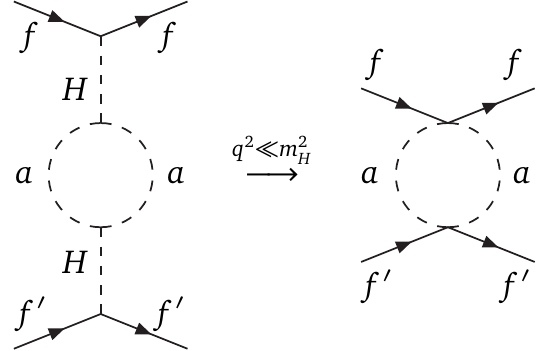}
  \end{minipage}%
\caption{Long-range force mediated by double neutrino exchange (left) and
  double (Pseudo-)Goldstone boson exchange. In both cases, the direct
  coupling to the fermions is by a heavy particle (\(Z\) or Higgs boson)
  leading to effective four-particle interactions.}
\label{fig:doubleparticleexch}
\par\medskip
\hrule
\end{figure}

The Standard Model already comprises a suppressed short-range
Yukawa-like potential mediated by the heavy electroweak vector bosons
or the scalar Higgs boson. According to the decoupling theorem, these
interactions should not have any effect on atomic or molecular scales
so that they can be safely ignored. However, there are claims in the
literature that an effective coupling mediated by heavy \(W\)\!, \(Z\)
or the Higgs boson leads to a measurable two-particle exchange of a
very light mediator. This two-fermion exchange may induce long-range
forces as pointed out in the literature~\cite{Feinberg:1968zz,
  Chang:1969rh, Hsu:1992tg, Grifols:1996fk, Ferrer:1998rw,
  Fichet:2017bng, Stadnik:2017yge, Thien:2019ayp, Ghosh:2019dmi,
  Bolton:2020xsm}, see Figure~\ref{fig:doubleparticleexch}.  For
instance, the case of an effective Fermi interaction with massless
neutrinos has been first discussed by Feinberg and Sucher in the late
1960s~\cite{Feinberg:1968zz} and was completed by Hsu and Sikivie in
the early 1990s~\cite{Hsu:1992tg}. Their work has been extended by
Grifols et al.~\cite{Grifols:1996fk} to the case of massive Dirac and
Majorana-type neutrinos of mass \( m_\nu \), yielding the long-range
potentials
\begin{subequations}\label{eq:neutrinopots}
  \begin{align}
    V_\mathrm{M}(r) & = \frac{G_\text{eff}^2 m_\nu^2}{8\pi^2r^3} \; K_2(2 m_\nu r) \stackrel{m_\nu r \gg 1}{\approx} \frac{G_\text{eff}}{16\pi^2 r^2} \sqrt{\frac{m_\nu^3}{\pi r^3}} \operatorname{e}^{-2m_\nu r}\,,
    \\
    V_\mathrm{D}(r) & = \frac{G_\mathrm{eff}^2m_\nu^3}{16\pi^3 r^2} \; K_3(2 m_\nu r) \stackrel{m_\nu r \gg 1}{\approx} \frac{G_\text{eff}}{32\pi^2} \sqrt{\frac{m_\nu^5}{\pi r^5}} \operatorname{e}^{-2m_\nu r}\,,
  \end{align}
\end{subequations}
with the modified Bessel functions \(K_n\). In the Standard Model
case, the effective coupling \( G_\text{eff} \) is given by the
Fermi constant \( G_\mathrm{F} \). Both potentials scale like \( \sim 1/r^5 \)
in the limit of vanishing neutrino masses or short distances,
reproducing the well-known result by Feinberg and Sucher~\cite{Feinberg:1968zz},
\begin{align}
    V(r) = \frac{G_\mathrm{eff}^2}{16 \pi^3 r^5}\,.
\end{align}
Due to the highly singular behaviour of the two-neutrino exchange
potential, one needs to be careful in the analysis. Naively, we expect
a quadratic divergence in our integrals from power-counting, matching
Stadnik's observation for hydrogen atoms in
Reference~\cite{Stadnik:2017yge}. Assuming a cut-off of the \( Z \)
boson mass \( M_\mathrm{Z} \) for the integrals to account for the
limited validity of the effective theory, Stadnik derives tight bounds
on the coupling \( G_\text{eff} \) close to the Standard Model value
for positronium. However, we doubt the adequateness of such a cut-off,
as remarked by other authors~\cite{Asaka:2018qfg,
  Costantino:2020bei}. The strong dependence of a result on the
arbitrary value of the cut-off parameter signals an incorrect
treatment of the ultraviolet divergences of the theory, bearing in
mind that already at the inverse Compton wavelength of the electron
much below \(M_\text{Z}\) the physics of the pure two-neutrino
potential changes.

\begin{figure}[tb]
  \centering
  \includegraphics[width=.3\textwidth]{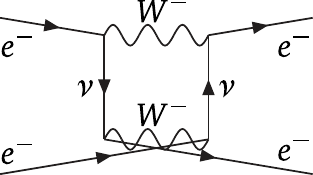}
  \caption{Additional contribution to the two-neutrino force induced
    by a \(W\)-box diagram.} \label{fig:Wbox} \par\medskip \hrule
\end{figure}

A correct treatment involving a proper matching in the tower of
effective theories at each scale in the problem is subtle and beyond
the scope of this analysis. In order to get an impression of the
magnitude of the neutrino exchange effects, we calculate the
short-range effect to the molecular levels caused by the exchange of a
\( W \) boson with mass \( M_\mathrm{W} \) as depicted in the diagram
in Figure~\ref{fig:Wbox}. Parametrically, the effects of both this
\(W\) box diagram and the neutrino exchange are of order
\(\order{G_\mathrm{F}^2 M_\mathrm{W}^2}\).  We are rather expecting a
further suppression, for instance due to small mixing factors in the
case of sterile neutrinos. Evaluating the box diagram, we derive the
effective low-energy potential
\begin{equation}
  V_{W\mathrm{box}}(\vb{r}) = \frac{4}{3}\pi G_\mathrm{F}^2 M_\mathrm{W}^2 \delta^{(3)}(\vb{r}) \,.
\end{equation}
The size of this effect is of \(\order{\SI{e-11}{\centi\meter^{-1}}}\)
and, hence, far below the current experimental sensitivity.  Due to
its smallness, the neutrino exchange is negligible if the Coulomb
force is present and one may rather expect an effect in cases where
the electromagnetic force is absent or screened like between neutral
atoms and molecules, as has been noted in
Reference~\cite{Feinberg:1968zz}.

The two-neutrino exchange has recently been studied in the context of
atomic parity violation~\cite{Arcadi:2019uif, Ghosh:2019dmi}. By using
higher angular momentum transitions, the authors are able to derive
limits from wave functions dropping rapidly for small distances, which
seems to be a suitable approach to deal with the divergence. However,
their analysis is missing a full systematic treatment of all matchings
at intermediate scales, which might have an influence on the
results. Nevertheless, the effect is far below the reach of current
and future experiments, similar to our estimate. The study of
Reference~\cite{Arcadi:2019uif} goes beyond the known physics
properties of electroweak physics and constrains new models extending
the Standard Model, especially with an extra Higgs doublet or a light
\(Z'\). A recent discussion of the full effective theory framework of
the long-range neutrino potential has been given
in~\cite{Bolton:2020xsm}.

There are further modifications of the intramolecular forces possible
due to Higgs and Goldstone boson exchange, where a long-range
potential arises in a similar manner as for
neutrinos~\cite{Ferrer:1998rw}, see
Figure~\ref{fig:doubleparticleexch}. In particular, a massive
(pseudo)scalar boson \( a \) of mass \( m \) interacting with the
Standard Model Higgs boson \( H \) via the Lagrangian \(
\Lag_\mathrm{Haa} = g_\mathrm{Haa} \: aaH \) yields the potential
\begin{equation}
  V_\text{PG} (r) = -\frac{G\, m}{8\pi^3 r^2} K_1(2m r) \,.
\end{equation}
Here, the coupling strength \( G \) is related to the Standard Model
Higgs-fermion interaction \( g_\mathrm{Hff} \) by~\cite{Ferrer:1998rw}
\begin{equation}
  G = \frac{g_\mathrm{Hff} g_\mathrm{Hf'f'} g_\mathrm{Haa}^2}{m_\mathrm{H}^4} \sim \SI{e-19}{\giga\electronvolt^{-4}} g_\mathrm{Haa}^2 \,.
\end{equation}
Although the potential reduces to a well-behaved \( 1/r^3 \)
functional form for small masses \( m \), the tiny prefactor renders
this process impossible to observe in molecular spectra.

\section{Conclusions} \label{sec:conclusion}

In the present work, we have performed the first extensive and
systematic study of New Physics effects on molecular spectra.
Starting from available codes which give precise \emph{ab initio}
predictions in the Standard Model for transitions in hydrogen-like
molecules, we treated a variety of New Physics potentials as
perturbations and derived constraints on the new forces from direct
measurements.

Molecular spectroscopy, as well as atomic spectroscopy, essentially
probes the Coulomb potential and Quantum Electrodynamics with high
accuracy. From our analysis we conclude that New Physics effects are
unlikely to be seen in spectroscopy at all. Spectroscopical observations
of sufficiently large deviations would be in conflict with indirect
constraints stemming from astrophysics and cosmology.  Nevertheless,
spectroscopy is a complementary and direct test of the Standard Model in
the laboratory and is an important tool in the context of New Physics
searches in this only indirectly excluded parameter region. Despite the
expectation that heavier nuclei as in tritiated molecules give stronger
constraints than hydrogen alone, we do not observe higher sensitivity in
the isotopologues compared to hydrogen.

We have found that constraints on new interactions between electrons
and nuclei from molecular spectroscopy are compatible with atomic
spectroscopy, but the latter derives more stringent bounds of up to
three orders of magnitude. The same is true for probes of the
nucleus--nucleus interaction from rovibrational spectroscopy compared
to direct neutron scattering. Furthermore, in the case of a modified
electron--electron coupling, molecular spectroscopy is competitive
with Helium spectroscopy, although there are stronger limits of
approximately two orders of magnitude on the coupling
$g_{\mathrm{NP}}$ available from measurements of the anomalous
magnetic moment of the electron.

As an advantage to other direct techniques molecular spectroscopy
allows to probe a plethora of New Physics interactions between
different types of particles in one single measurement---assuming that
only one type of interaction is present at the time.  Further
improvements in both theoretical treatment of hydrogen-like molecules
and experimental precision are going to yield stronger
constraints. Moreover, we have found relatively loose constraints for
a certain mass window of the mediator particle for some potentials.
In case of spin-dependent forces, polarized probes may help to improve
the exclusion limits.

Searches for a new long-range force mediated by an exchange of two
light particles like neutrinos are not promising in spectroscopy since
the expected effect is too small due to parametric
suppression. Furthermore, a strong cut-off dependence appears when
divergences in the theory are not properly taken into account so that
the sensitivity is misestimated.  A full treatment of the effective
theories at all scales down to very short distances is beyond the
scope of this paper. In any case, we do not expect an effect that is
going to be visible in the next generation experiments.

During the finalization of this work, we got aware of a new set of
measurements including more lines for T\(_2\), DT, and
HT~\cite{Lai:2020lcs}. This study shows very good agreement with the
theoretical prediction.

\singlespacing
{\small
\section*{Acknowledgements} \enlargethispage*{2em} 
We thank Magnus Schl\"osser for initiation of
this study, initial discussion and intermediate feedback, and Ulrich
Nierste for many helpful discussions and advices. Furthermore, we want
to acknowledge Sonia Rani and Aman Sardwal from the IIT
Bombay for their participation in the early stages of the
project.  WGH wants to thank Andreas Ringwald for helpful conversions
in the beginning of the project. Additionally, we thank Kirill
Melnikov for helpful advice on the proper treatment of the cut-off
with singular potentials, and we thank Alejandro Segarra for a
discussion of the two-neutrino exchange potentials. We thank Ulrich
Nierste for a careful reading of the manuscript and comments. ML
acknowledges the support by the Doctoral School ``Karlsruhe School of
Elementary and Astroparticle Physics: Science and Technology'' and the
collaborative research center TRR 257 ``Particle Physics Phenomenology
after the Higgs Discovery''. MT acknowledges the support of the
DFG-funded Research Training Group 1694, ``Elementary particle physics
at highest energy and precision''.
}

  \newpage
  \appendix
  \section{Experimental data}\label{app:Experimental data}

Here, we list all experimental data that were used in our analysis in Section~\ref{sec:results}.

\begin{table}[h!]
    \centering
    \caption{List of all measurements used in our analysis. The transitions \( (v_1, J_1) \rightarrow (v_2,J_2) \) are characterized by the vibrational quantum numbers \( v_i \) and the angular momentums \( J_i \) of the involved levels \( (v_i,J_i) \). All numbers are given in \si{\centi\meter^{-1}}.} 
    \begin{tabular}{ccS[table-format=5.8(2)]c}
        \toprule
        molecule & transition & {energy} & reference \\
        \midrule
        H$_2$ & $(3, 5)$ $\rightarrow$ $(0, 3)$ & 12559.74952 +- 0.00005 & \cite{PhysRevA.85.024501} \\
        H$_2$ & $(1, 0)$ $\rightarrow$ $(0, 0)$ & 4161.16636 +- 0.00015 & \cite{NIU201444} \\
        H$_2$ & $(1, 1)$ $\rightarrow$ $(0, 1)$ & 4155.25400 +- 0.00021 & \cite{NIU201444} \\
        H$_2$ & $(1, 2)$ $\rightarrow$ $(0, 2)$ & 4143.46553 +- 0.00015 & \cite{NIU201444} \\
        H$_2$ & $(11, 1)$ $\rightarrow$ $(0, 0)$ & 32937.7554 +- 0.0016 & \cite{Trivikram2016} \\
        H$_2$ & $(11, 3)$ $\rightarrow$ $(0, 0)$ & 33186.4791 +- 0.0016 & \cite{Trivikram2016} \\
        H$_2$ & $(11, 4)$ $\rightarrow$ $(0, 0)$ & 33380.1025 +- 0.0033  & \cite{Trivikram2016} \\
        H$_2$ & $(11, 5)$ $\rightarrow$ $(0, 0)$ & 33615.5371 +- 0.0018 & \cite{Trivikram2016} \\
        \midrule
        HD & $(1, 0)$ $\rightarrow$ $(0, 0)$ & 3632.16052 +- 0.00022 & \cite{NIU201444} \\
        HD & $(1, 1)$ $\rightarrow$ $(0, 1)$ & 3628.30450 +- 0.00022 & \cite{NIU201444} \\
        HD & $(2, 2)$ $\rightarrow$ $(0, 1)$ & 7241.84935087 +- 0.00000067  & \cite{PhysRevLett.120.153002} \\
        HD & $(2, 3)$ $\rightarrow$ $(0, 2)$ & 7306.48322250 +- 0.00000093 & \cite{PhysRevLett.120.153002} \\
        HD & $(2, 4)$ $\rightarrow$ $(0, 3)$ & 7361.90317335 +- 0.00000093 & \cite{PhysRevLett.120.153002} \\
        \midrule
        D$_2$ & $(1, 0)$ $\rightarrow$ $(0, 0)$ & 2993.61706 +- 0.00015 & \cite{NIU201444} \\
        D$_2$ & $(1, 1)$ $\rightarrow$ $(0, 1)$ & 2991.50706 +- 0.00015 & \cite{NIU201444} \\
        D$_2$ & $(1, 2)$ $\rightarrow$ $(0, 2)$ & 2987.29352 +- 0.00015 & \cite{NIU201444} \\
        D$_2$ & $(0, 2)$ $\rightarrow$ $(0, 0)$ & 179.068 +- 0.002 & \cite{doi:10.1002/jrs.5499} \\
        D$_2$ & $(0, 3)$ $\rightarrow$ $(0, 1)$ & 297.533 +- 0.003 & \cite{doi:10.1002/jrs.5499} \\
        D$_2$ & $(0, 4)$ $\rightarrow$ $(0, 2)$ & 414.648 +- 0.002 & \cite{doi:10.1002/jrs.5499} \\
        \midrule
        T$_2$ & $(1, 0)$ $\rightarrow$ $(0, 0)$ & 2464.5052 +- 0.0004 & \cite{Trivikram:2018} \\
        T$_2$ & $(1, 1)$ $\rightarrow$ $(0, 1)$ & 2463.3494 +- 0.0003 & \cite{Trivikram:2018} \\
        T$_2$ & $(1, 2)$ $\rightarrow$ $(0, 2)$ & 2461.0388 +- 0.0004 & \cite{Trivikram:2018} \\
	    T$_2$ & $(1, 3)$ $\rightarrow$ $(0, 3)$ & 2457.5803 +- 0.0004 & \cite{Trivikram:2018} \\
        T$_2$ & $(1, 4)$ $\rightarrow$ $(0, 4)$ & 2452.9817 +- 0.0004 & \cite{Trivikram:2018} \\
        T$_2$ & $(1, 5)$ $\rightarrow$ $(0, 5)$ & 2447.2510 +- 0.0004 & \cite{Trivikram:2018} \\
        \midrule
        DT & $(1, 0)$ $\rightarrow$ $(0, 0)$ & 2743.34171 +- 0.0004 & \cite{Lai2019DT} \\
        DT & $(1, 1)$ $\rightarrow$ $(0, 1)$ & 2741.73204 +- 0.0033 & \cite{Lai2019DT} \\
        DT & $(1, 2)$ $\rightarrow$ $(0, 2)$ & 2738.51659 +- 0.0004 & \cite{Lai2019DT} \\
        DT & $(1, 3)$ $\rightarrow$ $(0, 3)$ & 2733.70470 +- 0.0004 & \cite{Lai2019DT} \\
        DT & $(1, 4)$ $\rightarrow$ $(0, 4)$ & 2727.30734 +- 0.0004 & \cite{Lai2019DT} \\
        DT & $(1, 5)$ $\rightarrow$ $(0, 5)$ & 2719.34193 +- 0.0004 & \cite{Lai2019DT} \\
        \bottomrule
    \end{tabular}
\end{table}

\newpage

\begingroup
\let\secfnt\undefined
\newfont{\secfnt}{ptmb8t at 10pt}
\setstretch{.5}
\bibliographystyle{utcaps}
\bibliography{NewPhysTrit}

\providecommand{\href}[2]{#2}\begingroup\raggedright\begin{thebibliography}{10}

\bibitem{Knapen:2017xzo}
S.~Knapen, T.~Lin, and K.~M. Zurek, ``{Light Dark Matter: Models and
  Constraints}'', \href{http://dx.doi.org/10.1103/PhysRevD.96.115021}{{\em
  Phys. Rev.} D96 no.~11, (2017) 115021},
\href{http://arxiv.org/abs/1709.07882}{{\ttfamily arXiv:1709.07882 [hep-ph]}}.

\bibitem{Dror:2019onn}
J.~A. Dror, G.~Elor, and R.~Mcgehee, ``{Directly Detecting Signals from
  Absorption of Fermionic Dark Matter}'',
  \href{http://dx.doi.org/10.1103/PhysRevLett.124.181301}{{\em Phys. Rev.
  Lett.} 124 no.~18, (2020) 18},
  \href{http://arxiv.org/abs/1905.12635}{{\ttfamily arXiv:1905.12635
  [hep-ph]}}.

\bibitem{Dror:2019dib}
J.~A. Dror, G.~Elor, and R.~Mcgehee, ``{Absorption of Fermionic Dark Matter by
  Nuclear Targets}'', \href{http://dx.doi.org/10.1007/JHEP02(2020)134}{{\em
  JHEP} 02 (2020) 134}, \href{http://arxiv.org/abs/1908.10861}{{\ttfamily
  arXiv:1908.10861 [hep-ph]}}.

\bibitem{Arvanitaki:2017nhi}
A.~Arvanitaki, S.~Dimopoulos, and K.~Van~Tilburg, ``{Resonant absorption of
  bosonic dark matter in molecules}'',
  \href{http://dx.doi.org/10.1103/PhysRevX.8.041001}{{\em Phys. Rev.} X8 no.~4,
  (2018) 041001},
\href{http://arxiv.org/abs/1709.05354}{{\ttfamily arXiv:1709.05354 [hep-ph]}}.

\bibitem{Fichet:2017bng}
S.~Fichet, ``{Quantum Forces from Dark Matter and Where to Find Them}'',
  \href{http://dx.doi.org/10.1103/PhysRevLett.120.131801}{{\em Phys. Rev.
  Lett.} 120 no.~13, (2018) 131801},
\href{http://arxiv.org/abs/1705.10331}{{\ttfamily arXiv:1705.10331 [hep-ph]}}.

\bibitem{Brax:2017xho}
P.~Brax, S.~Fichet, and G.~Pignol, ``{Bounding Quantum Dark Forces}'',
  \href{http://dx.doi.org/10.1103/PhysRevD.97.115034}{{\em Phys. Rev.} D97
  no.~11, (2018) 115034},
\href{http://arxiv.org/abs/1710.00850}{{\ttfamily arXiv:1710.00850 [hep-ph]}}.

\bibitem{Kolos:1962}
W.~Kolos and L.~Wolniewicz, ``A complete non-relativistic treatment of the H2
  molecule'', {\em Physics Letters} 2 no.~5, (1962) 222--223.
  \url{https://doi.org/10.1016/0031-9163(62)90235-4}.

\bibitem{Kolos:1963}
W.~Ko\l{}os and L.~Wolniewicz, ``Nonadiabatic theory for diatomic molecules and
  its application to the hydrogen molecule'', {\em Reviews of Modern Physics}
  35 no.~3, (1963) 473--483. \url{https://doi.org/10.1103/RevModPhys.35.473}.

\bibitem{Kolos:1964adiab}
W.~Ko\l{}os and L.~Wolniewicz, ``Accurate Adiabatic Treatment of the Ground
  State of the Hydrogen Molecule'', {\em The Journal of chemical physics} 41
  no.~12, (1964) 3663--3673. \url{https://doi.org/10.1063/1.1725796}.

\bibitem{Kolos:1964vibro}
W.~Ko\l{}os and L.~Wolniewicz, ``Accurate computation of vibronic energies and
  of some expectation values for H2, D2, and T2'', {\em The Journal of chemical
  physics} 41 no.~12, (1964) 3674--3678.
  \url{https://doi.org/10.1063/1.1725797}.

\bibitem{Kolos:1968impr}
W.~Ko\l{}os and L.~Wolniewicz, ``Improved theoretical ground-state energy of
  the hydrogen molecule'', {\em The Journal of chemical physics} 49 no.~1,
  (1968) 404--410. \url{https://doi.org/10.1063/1.1669836}.

\bibitem{Pachucki2008}
K.~Pachucki and J.~Komasa, ``Nonadiabatic corrections to the wave function and
  energy'', \href{http://dx.doi.org/10.1063/1.2952517}{{\em Journal of Chemical
  Physics} 129 no.~3, (2008) }.
  \url{https://www.scopus.com/inward/record.uri?eid=2-s2.0-47849090977&doi=10.1063%2f1.2952517&partnerID=40&md5=e010a4eb52b57211614b28f17a58e5b3}.
  cited By 47.

\bibitem{Pachucki:2009xy}
K.~Pachucki and J.~Komasa, ``Nonadiabatic corrections to rovibrational levels
  of H2'', \href{http://dx.doi.org/10.1063/1.3114680}{{\em The Journal of
  Chemical Physics} 130 no.~16, (2009) 164113},
  \href{http://arxiv.org/abs/https://doi.org/10.1063/1.3114680}{{\ttfamily
  https://doi.org/10.1063/1.3114680}}. \url{https://doi.org/10.1063/1.3114680}.

\bibitem{Pachucki:2010pccp}
K.~Pachucki and J.~Komasa, ``Rovibrational levels of HD'',
  \href{http://dx.doi.org/10.1039/C0CP00209G}{{\em Phys. Chem. Chem. Phys.} 12
  (2010) 9188--9196}. \url{http://dx.doi.org/10.1039/C0CP00209G}.

\bibitem{Pachucki:2015xy}
K.~Pachucki and J.~Komasa, ``Leading order nonadiabatic corrections to
  rovibrational levels of H2, D2, and T2'',
  \href{http://dx.doi.org/10.1063/1.4927079}{{\em The Journal of Chemical
  Physics} 143 no.~3, (2015) 034111},
  \href{http://arxiv.org/abs/https://doi.org/10.1063/1.4927079}{{\ttfamily
  https://doi.org/10.1063/1.4927079}}. \url{https://doi.org/10.1063/1.4927079}.

\bibitem{Puchalski:2018pdf}
M.~Puchalski, J.~Komasa, P.~Czachorowski, and K.~Pachucki, ``{Nonadiabatic QED
  correction to the dissociation energy of the hydrogen molecule}'',
  \href{http://dx.doi.org/10.1103/PhysRevLett.122.103003}{{\em Phys. Rev.
  Lett.} 122 no.~10, (2019) 103003},
\href{http://arxiv.org/abs/1812.02980}{{\ttfamily arXiv:1812.02980
  [physics.atom-ph]}}.

\bibitem{Puchalski:2019prl}
M.~Puchalski, J.~Komasa, P.~Czachorowski, and K.~Pachucki, ``Nonadiabatic QED
  Correction to the Dissociation Energy of the Hydrogen Molecule'',
  \href{http://dx.doi.org/10.1103/PhysRevLett.122.103003}{{\em Phys. Rev.
  Lett.} 122 (Mar, 2019) 103003}.
  \url{https://link.aps.org/doi/10.1103/PhysRevLett.122.103003}.

\bibitem{Puchalski:2019pra}
M.~Puchalski, J.~Komasa, A.~Spyszkiewicz, and K.~Pachucki, ``Dissociation
  energy of molecular hydrogen isotopologues'',
  \href{http://dx.doi.org/10.1103/PhysRevA.100.020503}{{\em Phys. Rev. A} 100
  (Aug, 2019) 020503}.
  \url{https://link.aps.org/doi/10.1103/PhysRevA.100.020503}.

\bibitem{Puchalski:2016xy}
M.~Puchalski, J.~Komasa, P.~Czachorowski, and K.~Pachucki, ``Complete
  ${\ensuremath{\alpha}}^{6}\text{ }m$ Corrections to the Ground State of
  ${\mathrm{H}}_{2}$'',
  \href{http://dx.doi.org/10.1103/PhysRevLett.117.263002}{{\em Phys. Rev.
  Lett.} 117 (Dec, 2016) 263002}.
  \url{https://link.aps.org/doi/10.1103/PhysRevLett.117.263002}.

\bibitem{Puchalski:2017pra}
M.~Puchalski, J.~Komasa, and K.~Pachucki, ``Relativistic corrections for the
  ground electronic state of molecular hydrogen'',
  \href{http://dx.doi.org/10.1103/PhysRevA.95.052506}{{\em Phys. Rev. A} 95
  (May, 2017) 052506}.
  \url{https://link.aps.org/doi/10.1103/PhysRevA.95.052506}.

\bibitem{Puchalski:2018xy}
M.~Puchalski, A.~Spyszkiewicz, J.~Komasa, and K.~Pachucki, ``Nonadiabatic
  Relativistic Correction to the Dissociation Energy of ${\mathrm{H}}_{2}$,
  ${\mathrm{D}}_{2}$, and HD'',
  \href{http://dx.doi.org/10.1103/PhysRevLett.121.073001}{{\em Phys. Rev.
  Lett.} 121 (Aug, 2018) 073001}.
  \url{https://link.aps.org/doi/10.1103/PhysRevLett.121.073001}.

\bibitem{H2spectre}
P.~Czachorowski, J.~Komasa, G.~\L{}ach, K.~Pachucki, and M.~Puchalski,
  ``H2SPECTRE''. \url{https://www.fuw.edu.pl/~krp/codes.html};
  \url{https://qcg.home.amu.edu.pl/qcg/puclic_html/H2Spectre.html}.

\bibitem{czachothes}
P.~Czachorowski, {\em Relativistic nonadiabatic corrections to the ground state
  of molecular hydrogen}.
\newblock PhD thesis, University of Warsaw, 2019.

\bibitem{Komasa:2019}
J.~Komasa, M.~Puchalski, P.~Czachorowski, G.~\L{}ach, and K.~Pachucki,
  ``Rovibrational energy levels of the hydrogen molecule through nonadiabatic
  perturbation theory'',
  \href{http://dx.doi.org/10.1103/PhysRevA.100.032519}{{\em Phys. Rev. A} 100
  (Sep, 2019) 032519}.
  \url{https://link.aps.org/doi/10.1103/PhysRevA.100.032519}.

\bibitem{Salumbides:2013aga}
E.~J. Salumbides, J.~C.~J. Koelemeij, J.~Komasa, K.~Pachucki, K.~S.~E. Eikema,
  and W.~Ubachs, ``{Bounds on fifth forces from precision measurements on
  molecules}'', \href{http://dx.doi.org/10.1103/PhysRevD.87.112008}{{\em Phys.
  Rev.} D87 no.~11, (2013) 112008},
\href{http://arxiv.org/abs/1304.6560}{{\ttfamily arXiv:1304.6560
  [physics.atom-ph]}}.

\bibitem{Salumbides:2013dua}
E.~J. Salumbides, W.~Ubachs, and V.~I. Korobov, ``{Bounds on fifth forces at
  the sub-Angstrom length scale}'',
  \href{http://dx.doi.org/10.1016/j.jms.2014.04.003}{{\em J. Molec. Spectrosc.}
  300 (2014) 65},
\href{http://arxiv.org/abs/1308.1711}{{\ttfamily arXiv:1308.1711 [hep-ph]}}.

\bibitem{Salumbides:2015qwa}
E.~J. Salumbides, A.~N. Schellekens, B.~Gato-Rivera, and W.~Ubachs,
  ``{Constraints on extra dimensions from precision molecular spectroscopy}'',
  \href{http://dx.doi.org/10.1088/1367-2630/17/3/033015}{{\em New J. Phys.} 17
  no.~3, (2015) 033015},
\href{http://arxiv.org/abs/1502.02838}{{\ttfamily arXiv:1502.02838
  [physics.atom-ph]}}.

\bibitem{Jaeckel:2010xx}
J.~Jaeckel and S.~Roy, ``{Spectroscopy as a test of Coulomb's law: A Probe of
  the hidden sector}'',
  \href{http://dx.doi.org/10.1103/PhysRevD.82.125020}{{\em Phys. Rev.} D82
  (2010) 125020},
\href{http://arxiv.org/abs/1008.3536}{{\ttfamily arXiv:1008.3536 [hep-ph]}}.

\bibitem{Fadeev:2018rfl}
P.~Fadeev, Y.~V. Stadnik, F.~Ficek, M.~G. Kozlov, V.~V. Flambaum, and
  D.~Budker, ``{Revisiting spin-dependent forces mediated by new bosons:
  Potentials in the coordinate-space representation for macroscopic- and
  atomic-scale experiments}'',
  \href{http://dx.doi.org/10.1103/PhysRevA.99.022113}{{\em Phys. Rev.} A99
  no.~2, (2019) 022113},
\href{http://arxiv.org/abs/1810.10364}{{\ttfamily arXiv:1810.10364 [hep-ph]}}.

\bibitem{Costantino:2019ixl}
A.~Costantino, S.~Fichet, and P.~Tanedo, ``{Exotic Spin-Dependent Forces from a
  Hidden Sector}'', \href{http://dx.doi.org/10.1007/JHEP03(2020)148}{{\em JHEP}
  03 (2020) 148}, \href{http://arxiv.org/abs/1910.02972}{{\ttfamily
  arXiv:1910.02972 [hep-ph]}}.

\bibitem{Silveira:1985rk}
V.~Silveira and A.~Zee, ``{SCALAR PHANTOMS}'',
\href{http://dx.doi.org/10.1016/0370-2693(85)90624-0}{{\em Phys. Lett.} 161B
  (1985) 136--140}.

\bibitem{McDonald:1993ex}
J.~McDonald, ``{Gauge singlet scalars as cold dark matter}'',
  \href{http://dx.doi.org/10.1103/PhysRevD.50.3637}{{\em Phys. Rev.} D50 (1994)
  3637--3649},
\href{http://arxiv.org/abs/hep-ph/0702143}{{\ttfamily arXiv:hep-ph/0702143
  [HEP-PH]}}.

\bibitem{Moody:1984ba}
J.~E. Moody and F.~Wilczek, ``{New macroscopic forces?}'',
\href{http://dx.doi.org/10.1103/PhysRevD.30.130}{{\em Phys. Rev.} D30 (1984)
  130}.

\bibitem{Jaeckel:2010ni}
J.~Jaeckel and A.~Ringwald, ``{The Low-Energy Frontier of Particle Physics}'',
  \href{http://dx.doi.org/10.1146/annurev.nucl.012809.104433}{{\em Ann. Rev.
  Nucl. Part. Sci.} 60 (2010) 405--437},
\href{http://arxiv.org/abs/1002.0329}{{\ttfamily arXiv:1002.0329 [hep-ph]}}.

\bibitem{Holdom:1985ag}
B.~Holdom, ``{Two U(1)'s and Epsilon Charge Shifts}'',
\href{http://dx.doi.org/10.1016/0370-2693(86)91377-8}{{\em Phys. Lett.} 166B
  (1986) 196--198}.

\bibitem{Fayet:1990wx}
P.~Fayet, ``{Extra U(1)'s and New Forces}'',
\href{http://dx.doi.org/10.1016/0550-3213(90)90381-M}{{\em Nucl. Phys.} B347
  (1990) 743--768}.

\bibitem{Essig:2013lka}
R.~Essig {\em et~al.}, ``{Working Group Report: New Light Weakly Coupled
  Particles}'', in {\em {Proceedings, 2013 Community Summer Study on the Future
  of U.S. Particle Physics: Snowmass on the Mississippi (CSS2013): Minneapolis,
  MN, USA, July 29-August 6, 2013}}.
\newblock 2013.
\newblock
\href{http://arxiv.org/abs/1311.0029}{{\ttfamily arXiv:1311.0029 [hep-ph]}}.
\newblock

\bibitem{Delaunay:2017dku}
C.~Delaunay, C.~Frugiuele, E.~Fuchs, and Y.~Soreq, ``{Probing new
  spin-independent interactions through precision spectroscopy in atoms with
  few electrons}'', \href{http://dx.doi.org/10.1103/PhysRevD.96.115002}{{\em
  Phys. Rev.} D96 no.~11, (2017) 115002},
\href{http://arxiv.org/abs/1709.02817}{{\ttfamily arXiv:1709.02817 [hep-ph]}}.

\bibitem{Berengut:2017zuo}
J.~C. Berengut {\em et~al.}, ``{Probing New Long-Range Interactions by Isotope
  Shift Spectroscopy}'',
  \href{http://dx.doi.org/10.1103/PhysRevLett.120.091801}{{\em Phys. Rev.
  Lett.} 120 (2018) 091801},
\href{http://arxiv.org/abs/1704.05068}{{\ttfamily arXiv:1704.05068 [hep-ph]}}.

\bibitem{Jones:2019qny}
M.~P.~A. Jones, R.~M. Potvliege, and M.~Spannowsky, ``{Probing new physics
  using Rydberg states of atomic hydrogen}'',
\href{http://arxiv.org/abs/1909.09194}{{\ttfamily arXiv:1909.09194 [hep-ph]}}.

\bibitem{Kamiya:2015eva}
Y.~Kamiya, K.~Itagaki, M.~Tani, G.~N. Kim, and S.~Komamiya, ``{Constraints on
  New Gravitylike Forces in the Nanometer Range}'',
  \href{http://dx.doi.org/10.1103/PhysRevLett.114.161101}{{\em Phys. Rev.
  Lett.} 114 (2015) 161101},
\href{http://arxiv.org/abs/1504.02181}{{\ttfamily arXiv:1504.02181 [hep-ex]}}.

\bibitem{Brax:2010gp}
P.~Brax and C.~Burrage, ``{Atomic Precision Tests and Light Scalar
  Couplings}'', \href{http://dx.doi.org/10.1103/PhysRevD.83.035020}{{\em Phys.
  Rev.} D83 (2011) 035020},
\href{http://arxiv.org/abs/1010.5108}{{\ttfamily arXiv:1010.5108 [hep-ph]}}.

\bibitem{Villalba-Chavez:2018eql}
S.~Villalba-Chavez, A.~Golub, and C.~Muller, ``{Axion-modified photon
  propagator, Coulomb potential and Lamb-shift}'',
  \href{http://dx.doi.org/10.1103/PhysRevD.98.115008}{{\em Phys. Rev.} D98
  no.~11, (2018) 115008},
\href{http://arxiv.org/abs/1806.10940}{{\ttfamily arXiv:1806.10940 [hep-ph]}}.

\bibitem{Raffelt:1990yz}
G.~G. Raffelt, ``{Astrophysical methods to constrain axions and other novel
  particle phenomena}'',
  \href{http://dx.doi.org/10.1016/0370-1573(90)90054-6}{{\em Phys.\ Rept.} 198
  (1990) 1--113}.

\bibitem{Raffelt:1996wa}
G.~G. Raffelt, {\em {Stars as laboratories for fundamental physics}}.
\newblock Chicago, USA: Univ. Pr. (1996) 664 p, 1996.
\newblock
\url{http://wwwth.mpp.mpg.de/members/raffelt/mypapers/199613.pdf}.
\newblock

\bibitem{Raffelt:2006cw}
G.~G. Raffelt, ``{Astrophysical axion bounds}'',
  \href{http://dx.doi.org/10.1007/978-3-540-73518-2_3}{{\em Lect. Notes Phys.}
  741 (2008) 51--71},
\href{http://arxiv.org/abs/hep-ph/0611350}{{\ttfamily arXiv:hep-ph/0611350
  [hep-ph]}}.

\bibitem{Depta:2020wmr}
P.~F. Depta, M.~Hufnagel, and K.~Schmidt-Hoberg, ``{Robust cosmological
  constraints on axion-like particles}'',
\href{http://arxiv.org/abs/2002.08370}{{\ttfamily arXiv:2002.08370 [hep-ph]}}.

\bibitem{Carugno:1996uc}
G.~Carugno, Z.~Fontana, R.~Onofrio, and C.~Rizzo, ``{Limits on the existence of
  scalar interactions in the submillimeter range}'',
\href{http://dx.doi.org/10.1103/PhysRevD.55.6591}{{\em Phys. Rev.} D55 (1997)
  6591--6595}.

\bibitem{Wagner:2012ui}
T.~A. Wagner, S.~Schlamminger, J.~H. Gundlach, and E.~G. Adelberger,
  ``{Torsion-balance tests of the weak equivalence principle}'',
  \href{http://dx.doi.org/10.1088/0264-9381/29/18/184002}{{\em Class. Quant.
  Grav.} 29 (2012) 184002},
\href{http://arxiv.org/abs/1207.2442}{{\ttfamily arXiv:1207.2442 [gr-qc]}}.

\bibitem{Klimchitskaya:2015kxa}
G.~L. Klimchitskaya and V.~M. Mostepanenko, ``{Constraints on axion and
  corrections to Newtonian gravity from the Casimir effect}'',
  \href{http://dx.doi.org/10.1134/S0202289315010077}{{\em Grav. Cosmol.} 21
  no.~1, (2015) 1--12},
\href{http://arxiv.org/abs/1502.07647}{{\ttfamily arXiv:1502.07647 [hep-ph]}}.

\bibitem{Klimchitskaya:2017cnn}
G.~L. Klimchitskaya and V.~M. Mostepanenko, ``{Constraints on axionlike
  particles and non-Newtonian gravity from measuring the difference of Casimir
  forces}'', \href{http://dx.doi.org/10.1103/PhysRevD.95.123013}{{\em Phys.
  Rev.} D95 no.~12, (2017) 123013},
\href{http://arxiv.org/abs/1704.05892}{{\ttfamily arXiv:1704.05892 [hep-ph]}}.

\bibitem{Dupays:2006dp}
A.~Dupays, E.~Masso, J.~Redondo, and C.~Rizzo, ``{Light scalars coupled to
  photons and non-newtonian forces}'',
  \href{http://dx.doi.org/10.1103/PhysRevLett.98.131802}{{\em Phys. Rev. Lett.}
  98 (2007) 131802},
\href{http://arxiv.org/abs/hep-ph/0610286}{{\ttfamily arXiv:hep-ph/0610286
  [hep-ph]}}.

\bibitem{Dickenson:2013}
G.~D. Dickenson, M.~L. Niu, E.~J. Salumbides, J.~Komasa, K.~S.~E. Eikema,
  K.~Pachucki, and W.~Ubachs, ``Fundamental Vibration of Molecular Hydrogen'',
  \href{http://dx.doi.org/10.1103/PhysRevLett.110.193601}{{\em Phys. Rev.
  Lett.} 110 (May, 2013) 193601}.
  \url{https://link.aps.org/doi/10.1103/PhysRevLett.110.193601}.

\bibitem{NIU201444}
M.~Niu, E.~Salumbides, G.~Dickenson, K.~Eikema, and W.~Ubachs, ``Precision
  spectroscopy of the $X1\Sigma g+,v=0\rightarrow 1(J=0-2)$ rovibrational
  splittings in H2, HD and D2'',
  \href{http://dx.doi.org/https://doi.org/10.1016/j.jms.2014.03.011}{{\em
  Journal of Molecular Spectroscopy} 300 (2014) 44 -- 54}.
  \url{http://www.sciencedirect.com/science/article/pii/S0022285214000630}.
  Spectroscopic Tests of Fundamental Physics.

\bibitem{PhysRevLett.120.153002}
F.~M.~J. Cozijn, P.~Dupr\'e, E.~J. Salumbides, K.~S.~E. Eikema, and W.~Ubachs,
  ``Sub-Doppler Frequency Metrology in HD for Tests of Fundamental Physics'',
  \href{http://dx.doi.org/10.1103/PhysRevLett.120.153002}{{\em Phys. Rev.
  Lett.} 120 (Apr, 2018) 153002}.
  \url{https://link.aps.org/doi/10.1103/PhysRevLett.120.153002}.

\bibitem{Schlosser_2017}
M.~Schl\"osser, X.~Zhao, M.~Trivikram, W.~Ubachs, and E.~J. Salumbides,
  ``{CARS} spectroscopy of the \((v=0 \to 1)\) band in \(\mathrm{T}_2 \)'',
  \href{http://dx.doi.org/10.1088/1361-6455/aa8d80}{{\em Journal of Physics B:
  Atomic, Molecular and Optical Physics} 50 no.~21, (Oct, 2017) 214004}.
  \url{https://doi.org/10.1088%2F1361-6455%2Faa8d80}.

\bibitem{Trivikram:2018}
T.~M. Trivikram, M.~Schl\"osser, W.~Ubachs, and E.~J. Salumbides,
  ``Relativistic and QED Effects in the Fundamental Vibration of
  ${\mathrm{T}}_{2}$'',
  \href{http://dx.doi.org/10.1103/PhysRevLett.120.163002}{{\em Phys. Rev.
  Lett.} 120 (Apr, 2018) 163002}.
  \url{https://link.aps.org/doi/10.1103/PhysRevLett.120.163002}.

\bibitem{Lai2019DT}
K.-F. Lai, P.~Czachorowski, M.~Schl\"osser, M.~Puchalski, J.~Komasa,
  K.~Pachucki, W.~Ubachs, and E.~J. Salumbides, ``Precision tests of
  nonadiabatic perturbation theory with measurements on the DT molecule'',
  \href{http://dx.doi.org/10.1103/PhysRevResearch.1.033124}{{\em Phys. Rev.
  Research} 1 (Nov, 2019) 033124}.
  \url{https://link.aps.org/doi/10.1103/PhysRevResearch.1.033124}.

\bibitem{Heitler:1927}
W.~Heitler and F.~London, ``Wechselwirkung neutraler Atome und hom{\"o}opolare
  Bindung nach der Quantenmechanik'',
  \href{http://dx.doi.org/10.1007/BF01397394}{{\em Zeitschrift f{\"u}r Physik}
  44 no.~6, (Jun, 1927) 455--472}. \url{https://doi.org/10.1007/BF01397394}.

\bibitem{Born:1927}
M.~Born and R.~Oppenheimer, ``Zur Quantentheorie der Molekeln'',
  \href{http://dx.doi.org/10.1002/andp.19273892002}{{\em Annalen der Physik}
  389 no.~20, 457--484}.
  \url{https://onlinelibrary.wiley.com/doi/abs/10.1002/andp.19273892002}.

\bibitem{Born:1951}
M.~Born, {\em Kopplung der Elektronen- und Kernbewegung in Molekeln und
  Kristallen}.
\newblock Nachrichten der Akademie der Wissenschaften in G\"ottingen,
  Mathematisch-Physikalische Klasse: 2a, Mathematisch-Physikalisch-Chemische
  Abteilung; 1951, 6. Vandenhoeck \& Ruprecht, G\"ottingen, 1951.

\bibitem{doi:10.1063/1.1749252}
H.~M. James and A.~S. Coolidge, ``The Ground State of the Hydrogen Molecule'',
  \href{http://dx.doi.org/10.1063/1.1749252}{{\em The Journal of Chemical
  Physics} 1 no.~12, (1933) 825--835},
  \href{http://arxiv.org/abs/https://doi.org/10.1063/1.1749252}{{\ttfamily
  https://doi.org/10.1063/1.1749252}}. \url{https://doi.org/10.1063/1.1749252}.

\bibitem{PhysRevA.82.032509}
K.~Pachucki, ``Born-Oppenheimer potential for ${\mathrm{H}}_{2}$'',
  \href{http://dx.doi.org/10.1103/PhysRevA.82.032509}{{\em Phys. Rev. A} 82
  (Sep, 2010) 032509}.
  \url{https://link.aps.org/doi/10.1103/PhysRevA.82.032509}.

\bibitem{Czachorowski:2018xvk}
P.~Czachorowski, M.~Puchalski, J.~Komasa, and K.~Pachucki, ``{Nonadiabatic
  relativistic correction in H2 , D2 , and HD}'',
  \href{http://dx.doi.org/10.1103/PhysRevA.98.052506}{{\em Phys. Rev.} A98
  no.~5, (2018) 052506},
\href{http://arxiv.org/abs/1810.02604}{{\ttfamily arXiv:1810.02604
  [physics.atom-ph]}}.

\bibitem{Puchalski:2016ibj}
M.~Puchalski, J.~Komasa, P.~Czachorowski, and K.~Pachucki, ``{Complete
  $\alpha^6\,m$ corrections to the ground state of H$_2$}'',
  \href{http://dx.doi.org/10.1103/PhysRevLett.117.263002}{{\em Phys. Rev.
  Lett.} 117 no.~26, (2016) 263002},
\href{http://arxiv.org/abs/1608.07081}{{\ttfamily arXiv:1608.07081
  [physics.chem-ph]}}.

\bibitem{Pachucki:2016H2Solv}
K.~Pachucki, M.~Zientkiewicz, and V.~Yerokhin, ``H2SOLV: Fortran solver for
  diatomic molecules in explicitly correlated exponential basis'',
  \href{http://dx.doi.org/10.1016/j.cpc.2016.07.024}{{\em Computer Physics
  Communications} 208 (07, 2016) }.

\bibitem{BIRABEN2019671}
F.~Biraben, ``The first decades of Doppler-free two-photon spectroscopy'',
  \href{http://dx.doi.org/https://doi.org/10.1016/j.crhy.2019.04.003}{{\em
  Comptes Rendus Physique} 20 no.~7, (2019) 671 -- 681}.
  \url{http://www.sciencedirect.com/science/article/pii/S1631070519300283}. La
  science en mouvement 2 : de 1940 aux premi\`eres ann\'ees 1980 \^a Avanc\'ees
  en physique.

\bibitem{doi:10.1002/jrs.5499}
R.~Z. Mart\'{i}nez, D.~Bermejo, P.~Wcis\l{}o, and F.~Thibault, ``Accurate
  wavenumber measurements for the S0(0), S0(1), and S0(2) pure rotational Raman
  lines of D2'', \href{http://dx.doi.org/10.1002/jrs.5499}{{\em Journal of
  Raman Spectroscopy} 50 no.~1, 127--129}.
  \url{https://onlinelibrary.wiley.com/doi/abs/10.1002/jrs.5499}.

\bibitem{Feinberg:1968zz}
G.~Feinberg and J.~Sucher, ``{Long-Range Forces from Neutrino-Pair Exchange}'',
\href{http://dx.doi.org/10.1103/PhysRev.166.1638}{{\em Phys. Rev.} 166 (1968)
  1638--1644}.

\bibitem{Peccei:1977hh}
R.~D. Peccei and H.~R. Quinn, ``{CP Conservation in the Presence of
  Instantons}'',
\href{http://dx.doi.org/10.1103/PhysRevLett.38.1440}{{\em Phys. Rev. Lett.} 38
  (1977) 1440--1443}.

\bibitem{Peccei:1977ur}
R.~D. Peccei and H.~R. Quinn, ``{Constraints Imposed by CP Conservation in the
  Presence of Instantons}'',
\href{http://dx.doi.org/10.1103/PhysRevD.16.1791}{{\em Phys. Rev.} D16 (1977)
  1791--1797}.

\bibitem{Wilczek:1977pj}
F.~Wilczek, ``{Problem of Strong $P$ and $T$ Invariance in the Presence of
  Instantons}'',
\href{http://dx.doi.org/10.1103/PhysRevLett.40.279}{{\em Phys. Rev. Lett.} 40
  (1978) 279--282}.

\bibitem{Weinberg:1977ma}
S.~Weinberg, ``{A New Light Boson?}'',
\href{http://dx.doi.org/10.1103/PhysRevLett.40.223}{{\em Phys. Rev. Lett.} 40
  (1978) 223--226}.

\bibitem{Preskill:1982cy}
J.~Preskill, M.~B. Wise, and F.~Wilczek, ``{Cosmology of the Invisible
  Axion}'',
\href{http://dx.doi.org/10.1016/0370-2693(83)90637-8}{{\em Phys. Lett.} B120
  (1983) 127--132}.

\bibitem{Chikashige:1980ui}
Y.~Chikashige, R.~N. Mohapatra, and R.~D. Peccei, ``{Are There Real Goldstone
  Bosons Associated with Broken Lepton Number?}'',
\href{http://dx.doi.org/10.1016/0370-2693(81)90011-3}{{\em Phys. Lett.} 98B
  (1981) 265--268}.

\bibitem{Stueckelberg:1900zz}
E.~C.~G. Stueckelberg, ``{Interaction energy in electrodynamics and in the
  field theory of nuclear forces}'',
\href{http://dx.doi.org/10.5169/seals-110852}{{\em Helv. Phys. Acta} 11 (1938)
  225--244}.

\bibitem{Kors:2004dx}
B.~Kors and P.~Nath, ``{A Stueckelberg extension of the standard model}'',
  \href{http://dx.doi.org/10.1016/j.physletb.2004.02.051}{{\em Phys. Lett.}
  B586 (2004) 366--372},
\href{http://arxiv.org/abs/hep-ph/0402047}{{\ttfamily arXiv:hep-ph/0402047
  [hep-ph]}}.

\bibitem{Chang:1969rh}
S.-J. Chang and R.~Rajaraman, ``{Long-range corrections to the coulomb
  potential and their implications about weak interactions}'',
\href{http://dx.doi.org/10.1103/PhysRev.183.1442}{{\em Phys. Rev.} 183 (1969)
  1442--1445}.

\bibitem{Hsu:1992tg}
S.~D.~H. Hsu and P.~Sikivie, ``{Long range forces from two neutrino exchange
  revisited}'', \href{http://dx.doi.org/10.1103/PhysRevD.49.4951}{{\em Phys.
  Rev.} D49 (1994) 4951--4953},
\href{http://arxiv.org/abs/hep-ph/9211301}{{\ttfamily arXiv:hep-ph/9211301
  [hep-ph]}}.

\bibitem{Grifols:1996fk}
J.~A. Grifols, E.~Masso, and R.~Toldra, ``{Majorana neutrinos and long range
  forces}'', \href{http://dx.doi.org/10.1016/S0370-2693(96)01304-4}{{\em Phys.
  Lett.} B389 (1996) 563--565},
\href{http://arxiv.org/abs/hep-ph/9606377}{{\ttfamily arXiv:hep-ph/9606377
  [hep-ph]}}.

\bibitem{Ferrer:1998rw}
F.~Ferrer and M.~Nowakowski, ``{Higgs and Goldstone bosons mediated long range
  forces}'', \href{http://dx.doi.org/10.1103/PhysRevD.59.075009}{{\em Phys.
  Rev.} D59 (1999) 075009},
\href{http://arxiv.org/abs/hep-ph/9810550}{{\ttfamily arXiv:hep-ph/9810550
  [hep-ph]}}.

\bibitem{Stadnik:2017yge}
Y.~V. Stadnik, ``{Probing Long-Range Neutrino-Mediated Forces with Atomic and
  Nuclear Spectroscopy}'',
  \href{http://dx.doi.org/10.1103/PhysRevLett.120.223202}{{\em Phys. Rev.
  Lett.} 120 no.~22, (2018) 223202},
\href{http://arxiv.org/abs/1711.03700}{{\ttfamily arXiv:1711.03700
  [physics.atom-ph]}}.

\bibitem{Thien:2019ayp}
Q.~Le~Thien and D.~E. Krause, ``Spin-Independent Two-Neutrino Exchange
  Potential with Mixing and $CP$-Violation'',
  \href{http://dx.doi.org/10.1103/PhysRevD.99.116006}{{\em Phys.Rev.D} 99
  no.~11, (2019) 116006}, \href{http://arxiv.org/abs/1901.05345}{{\ttfamily
  arXiv:1901.05345 [hep-ph]}}.

\bibitem{Ghosh:2019dmi}
M.~Ghosh, Y.~Grossman, and W.~Tangarife, ``{Probing the two-neutrino exchange
  force using atomic parity violation}'',
\href{http://arxiv.org/abs/1912.09444}{{\ttfamily arXiv:1912.09444 [hep-ph]}}.

\bibitem{Bolton:2020xsm}
P.~D. Bolton, F.~F. Deppisch, and C.~Hati, ``{Probing New Physics with
  Long-Range Neutrino Interactions: An Effective Field Theory Approach}'',
  \href{http://arxiv.org/abs/2004.08328}{{\ttfamily arXiv:2004.08328
  [hep-ph]}}.

\bibitem{Asaka:2018qfg}
T.~Asaka, M.~Tanaka, K.~Tsumura, and M.~Yoshimura, ``{Precision electroweak
  shift of muonium hyperfine splitting}'',
  \href{http://arxiv.org/abs/1810.05429}{{\ttfamily arXiv:1810.05429
  [hep-ph]}}.

\bibitem{Costantino:2020bei}
A.~Costantino and S.~Fichet, ``{The Neutrino Casimir Force}'',
  \href{http://arxiv.org/abs/2003.11032}{{\ttfamily arXiv:2003.11032
  [hep-ph]}}.

\bibitem{Arcadi:2019uif}
G.~Arcadi, M.~Lindner, J.~Martins, and F.~S. Queiroz, ``{New Physics Probes:
  Atomic Parity Violation, Polarized Electron Scattering and Neutrino-Nucleus
  Coherent Scattering}'',
\href{http://arxiv.org/abs/1906.04755}{{\ttfamily arXiv:1906.04755 [hep-ph]}}.

\bibitem{Lai:2020lcs}
K.-F. Lai, V.~Hermann, T.~Trivikram, M.~Diouf, M.~Schlösser, W.~Ubachs, and
  E.~Salumbides, ``{Precision measurement of the fundamental vibrational
  frequencies of tritium-bearing hydrogen molecules: T$_2$, DT, HT}'',
  \href{http://arxiv.org/abs/2003.11060}{{\ttfamily arXiv:2003.11060
  [physics.chem-ph]}}.

\bibitem{PhysRevA.85.024501}
C.-F. Cheng, Y.~R. Sun, H.~Pan, J.~Wang, A.-W. Liu, A.~Campargue, and S.-M. Hu,
  ``Electric-quadrupole transition of H${}_{2}$ determined to
  ${10}^{\ensuremath{-}9}$ precision'',
  \href{http://dx.doi.org/10.1103/PhysRevA.85.024501}{{\em Phys. Rev. A} 85
  (Feb, 2012) 024501}.
  \url{https://link.aps.org/doi/10.1103/PhysRevA.85.024501}.

\bibitem{Trivikram2016}
T.~M. Trivikram, M.~L. Niu, P.~Wcislo, W.~Ubachs, and E.~J. Salumbides,
  ``Precision measurements and test of molecular theory in highly excited
  vibrational states of H2 ($v=11$)'',
  \href{http://dx.doi.org/10.1007/s00340-016-6570-1}{{\em Applied Physics B}
  122 no.~12, (2016) 294}. \url{https://doi.org/10.1007/s00340-016-6570-1}.

\end{thebibliography}\endgroup
\endgroup

\end{document}